\documentclass[10pt,aps,prx,twocolumn,superscriptaddress,showpacs]{revtex4-2}
\usepackage{graphicx} 
\usepackage{mathtools,amssymb,amsthm}
\usepackage{color,xcolor}
\usepackage{bm}

\usepackage{tikz}
\usetikzlibrary{arrows.meta, positioning}
\usetikzlibrary{calc,arrows.meta}
\usepackage{verbatim}
\usepackage[utf8]{inputenc}

\begin{document}

\title{ A Full-Spectrum Theory of Moiré Superlattices}
\author{Tong Liu}
\affiliation{ Shenzhen Institute for Quantum Science and Enginneering, Southern University of Science and Technology, 518055, Shenzhen, P. R. China}
\affiliation{ Department of Physics, Hong Kong University of Science and Technology, Hong Kong }

\author{X. R. Wang}
\email{phxwan@cuhk.edu.cn}
\affiliation{ School of Science and Engineering, Chinese University of Hong Kong (Shenzhen), Shenzhen, 51817, China }

\author{Jiansheng Wu}
\email{jwu@sustech.edu.cn} 
\affiliation{ Shenzhen Institute for Quantum Science and Enginneering, Southern University of Science and Technology, 518055, Shenzhen, P. R. China}


\email{phxwan@cuhk.edu.cn (X.W.) jwu@sustech.edu.cn (J. W.)}

\date{August 2026}
\begin{abstract}
Conventional theoretical descriptions of moiré superlattices, epitomized by the Bistritzer–MacDonald continuum model, are fundamentally constrained by low-energy Dirac cone approximations and first-shell interlayer coupling truncations. While successful for small-angle twisted bilayer graphene near charge neutrality, these frameworks fail to provide universal analytical solutions for strained lattices, high-energy band behaviors, or arbitrary commensurate geometries. Here, we establish a unified, approximation-free analytical formalism for generalized moiré systems based on three pillars: (i) a rigorous two-dimensional-to-one-dimensional supercell cyclic mapping, (ii) exact block-circulant diagonalization of intralayer Hamiltonians, and (iii) quasi-degenerate Löwdin perturbation theory centered on arbitrary energy degeneracy points across the full spectrum. Unlike traditional schemes anchored to the Dirac point, our approach enables systematic perturbative expansions around any band crossing, unifying the characterization of low-energy flat bands with high-energy degeneracies. We demonstrate that this framework perfectly recovers the Bistritzer–MacDonald model in the low-energy limit while extending its reach to heterostrained systems and high-energy regimes. By incorporating high-order interlayer coupling shells, we reveal an intrinsic scale self-similarity in moiré lattices and predict the emergence of higher-order magic angles and tunable flat bands driven by secondary degeneracies. This work eliminates the long-standing lack of universal analytical tools for diverse moiré platforms, providing a full-spectrum theoretical foundation for band engineering and the exploration of exotic correlated and topological states in van der Waals heterostructures.
\end{abstract}
\maketitle

\section{Introduction}
Moir\'e superlattice is the large-scale periodic pattern that emerges when two periodic structures with slightly different periods, or one rotated relative to the other, are overlaid. The beating between the two lattices produces a new periodicity whose length is much larger than either original period. The most prominent examples are twist-engineered moir\'e systems, such as twisted bilayer graphene, twisted hexagonal boron nitride~\cite{Dean2013,Ponomarenko2013}, and twisted transition metal dichalcogenide heterostructures~\cite{Wu2018,Naik2018}. Strain-modified two-dimensional lattices, especially uniaxially and biaxially heterostrained bilayer graphene~\cite{Levy2010,Guinea2010,Naumis2017}, constitute another important branch of composite lattice systems. In addition, artificially designed quasiperiodic lattices (e.g., 
Fibonacci lattice configurations)~\cite{Shechtman1984,Levine1984,Kohmoto1983,Kraus2012}, traditional semiconductor compositional superlattices (GaAs/AlGaAs)~\cite{Esaki1970}, synthetic perovskite moir\'e superlattices~\cite{Tisdale2024}, artificial kagome lattices~\cite{Gomes2012}, and optical cold-atom lattice systems~\cite{Bloch2008,Cooper2019,Tarruell2012} also fall into this category. All these composite lattice platforms have emerged as core research focuses, and the exploration of their unique band structures and exotic electronic behaviors is essential for revealing novel topological characteristics and strongly correlated quantum states.

The standard theoretical paradigm for describing flat bands in twisted bilayer graphene~\cite{Cao2018a,Cao2018b} originates from the low-energy physical properties of pristine monolayer graphene. Traditional continuum modeling starts with the low-energy Dirac Hamiltonian of single-layer graphene~\cite{CastroNeto2009}, adopts continuous low-energy dispersion relations to characterize electronic behaviors of two graphene layers, and expands the interlayer hybridization interaction based on the first set of moir\'e reciprocal lattice vectors~\cite{Bistritzer2011,LopesSantos2007,LopesSantos2012}. This conventional construction is strictly anchored to low-energy Dirac cone features, retaining only the band degeneracy at the K and K$'$ high-symmetry points and focusing exclusively on the two electronic bands near charge neutrality, which fundamentally restricts its applicability to low-energy regimes. Early tight-binding simulations and first-principles theoretical analyses of slightly twisted bilayer graphene have successfully predicted the emergence of moir\'e flat bands~\cite{SuarezMorell2010,Trambly2010,Mele2010,Shallcross2010,Li2010,Brihuega2012}. Subsequent improvements to this classic framework have evolved along two mainstream directions. One branch develops multi-band wave vector-dependent perturbation schemes to expand the plane-wave basis set while still adhering to the low-energy Dirac cone approximation~\cite{MoonKoshino2012,Koshino2018}. The other branch establishes first-principles-based perturbation models by incorporating optimized tight-binding or density functional theory parameters to improve the quantitative accuracy of flat band calculations~\cite{Carr2019,FangKaxiras2016}, while exact analytical solutions within these frameworks are only achievable under the ideal chiral limit~\cite{Tarnopolsky2019}. Nevertheless, existing low-energy band descriptions can only explain the correlated quantum phenomena of magic-angle moir\'e superlattices~\cite{KangVafek2018,Po2018,Andrei2020,Balents2020,Kennes2021,Cao2016,Kim2017,Yoo2019,Sharpe2019,Serlin2019}, lacking universality for general composite lattice systems.

To break the inherent limitations of conventional low-energy-centered modeling strategies, the present work develops an inverse construction paradigm for moir\'e system modeling. Unlike traditional methods that start from empirical low-energy physical assumptions and conduct incremental optimization, 
our approach begins with the complete supercell tight-binding Hamiltonian and implements systematic and rigorous dimensional reduction and simplification. We first perform block diagonalization on the intralayer Hamiltonian with block-circulant structural characteristics via discrete Fourier transform, which enables accurate identification and isolation of arbitrary quasi-degenerate band states in the full energy spectrum. The interlayer coupling effect is then projected onto the isolated degenerate subspace through quasi-degenerate L\"owdin perturbation theory~\cite{Lowdin1951}. This novel theoretical framework possesses three prominent advantages over conventional methods. 
First, it completely abandons the low-energy approximation adopted by traditional models, thus covering the full energy spectrum~\cite{Liu2019}. Second, all commensurate moir\'e lattice geometries are processed precisely through a two-dimensional-to-one-dimensional supercell mapping technique based on supercell cyclic structures and modular arithmetic, realizing accurate discrete momentum transfer calculations beyond the small-angle approximation range. 
Third, high-order interlayer coupling shells are naturally derived from fundamental momentum conservation constraints without artificial empirical hypothesis. 
We verify that the secondary coupling shell exhibits identical phase characteristics to the primary shell, revealing the intrinsic scale self-similarity of moir\'e systems and providing a physical foundation for momentum rescaling modeling schemes~\cite{Wang2021,Navarro2022}.

Despite the superior universality and full-energy-spectrum applicability of the proposed framework, this generalized formalism also comes with three nontrivial computational and physical costs. 
First, the gauge condition for block diagonalization requires explicit artificial fixation, since all spectral physical quantities are gauge-invariant, while the effective interlayer coupling matrix lacks gauge invariance. 
Second, the accurate geometric description relying on modular inverse operation introduces additional algebraic computational overhead that does not exist in traditional simplified continuum models. 
Importantly, the proposed framework exhibits excellent self-consistency and compatibility: it can perfectly degenerate into the classic Bistritzer--MacDonald model under the low-energy and first-shell coupling limit. Beyond this limit, our method provides a unified and versatile theoretical platform for high-energy band characterization, arbitrary commensurate moir\'e geometries, real-space quasicrystalline structures, and various types of band degeneracy, with twisted bilayer graphene merely serving as a typical application scenario.

\section{A General Model for Moir\'e structure }
\begin{figure}[ht]
\centering
\includegraphics[width=\linewidth]{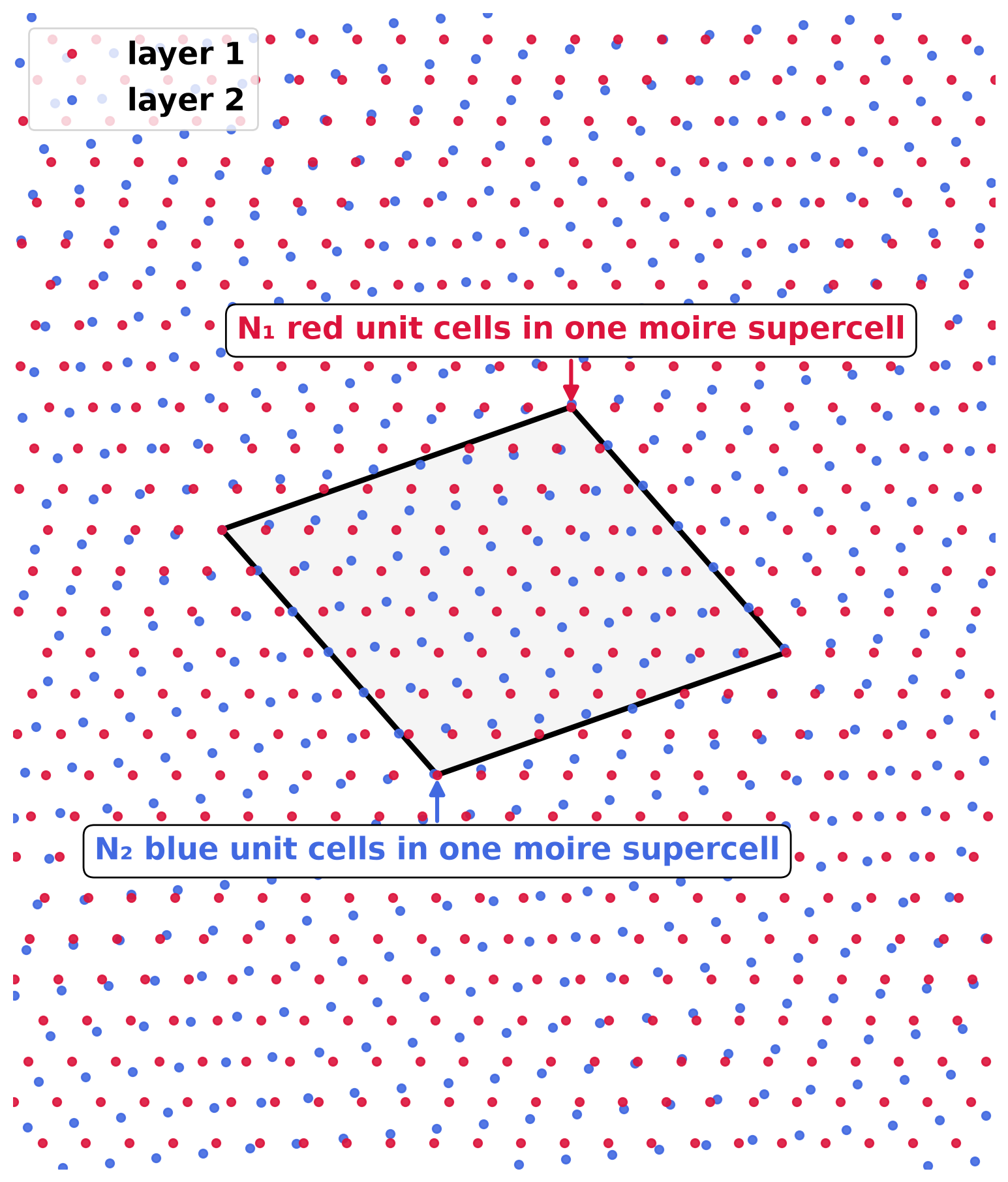}
\caption{Here red and blue points stand for unit cell in the two layers $1$ and $2$. The black box is the moir\'e superlattice structure, which contains $N_{1}$ unit cells from layer 1 and $N_{2}$ unit cells from layer 2. This determines the size of $H_{1}$ and $H_{2}$, shown as Eq.~(\ref{eq:bcmatrix}).}
\label{fig: moire}
\end{figure}

In this work, we target a general class of Moir\'e structure. 
In such Moir\'e superlattice system, the first and second layers 
for a superlattice struture, and each layer contain finite numbers of primitive unit cells. One supercell contain $N_i\ (i=1,2)$ unit cell of layer $i$ as shown in Fig.(~\ref{fig: moire}). If these two layers don't form a superlattice structure, $N_i$ is set to infinity.

Our theoretical framework is built upon two fundamental physical assumptions: the intrinsic lattice structure and complete tight-binding parameters of each individual constituent layer are fully known, and the interlayer atomic hopping amplitudes between the two layers can be quantitatively determined by the interlayer vertical distance, in-plane relative atomic displacement, and specific atomic species of the bilayer system. Some studies~\cite{Movassagh2017,Bolten2012} have shown that the Hamiltonian of a periodic crystalline superlattice in tight-binding model possesses a block-circulant structure. The Hamiltonian of such a system consists of intralayer and interlayer  terms and can be given by:
\begin{eqnarray}
H^{M}=\begin{bmatrix}
H_{1} & T_{12}\\
T_{12}^{\dagger} & H_{2}
\end{bmatrix}
\label{eq:gmatrix}
\end{eqnarray}
In this case $H_{i}$ is the Hamiltonian of $i$-layer lattice, which is a $N_{i}\times n_{i}$-th order square block-circulant matrix, where $n_{i}$ is the number of atoms in one unit cell in $i$-th layer. $T_{12}$ stands for interlayer hopping terms. In this case, we can write $H_{i}$ in this form:
\begin{eqnarray}
H_{i}= \begin{bmatrix}
a_{1} & a_{2} & a_{3} & ... & a_{N_{i}-1} & a_{N_{i}} \\
a_{N_{i}} & a_{1} & a_{2} & ... & a_{N_{i}-2} & a_{N_{i}-1} \\
a_{N_{i}-1} & a_{N_{i}} & a_{1} & ... & a_{N_{i}-3} & a_{N_{i}-2} \\
... & ... & ... & ... & ... & ... \\
a_{3} & a_{4} & a_{5} & ... & a_{1} & a_{2} \\
a_{2} & a_{3} & a_{4} & ... & a_{N_{i}} & a_{1} \\
\end{bmatrix}
\label{eq:bcmatrix}
\end{eqnarray}
Here each $a_{j}$ can be a $n_{i}$-th order Hermitian matrix which represents the Hamiltonian of $j$-th unit cell in this layer.Since $H_i$ is Hermitian, we have $a_j=a_{N_i+2-j}^\dag$.
Why $H_i$ can be written as block-circulant matrix is due to following fact. Crucially, with a proper choice of supercell basis vectors, all primitive cells contained within the two-dimensional supercell can be reordered into a one-dimensional periodic chain arrangement (detailed proof can be found in Section (IV)). Such structural rearrangement forms a closed cyclic periodic geometry for the supercell lattice sites, which fundamentally endows the intralayer Hamiltonian of each individual layer with a strict block-circulant structure. This intrinsic cyclic characteristic guarantees the feasibility of exact diagonalization via discrete Fourier transform and constitutes the core structural foundation for our subsequent systematic model reduction. 
The eigenvector of the matrix $H^{M}$ is denoted as $\Psi^{M}=(\Psi_{1}, \Psi_{2})$, ($\Psi_{i}=(C^{i}_{1}, C^{i}_{2},...,C^{i}_{N_{i}})$) and the eigenvalue is $E=\Psi^{M \dagger}H^{M}\Psi^{M}$:
\begin{eqnarray}
E=\sum_{i}\Psi_{i}^{\dagger}H_{i}\Psi_{i}+\Psi_{1}^{\dagger}T_{12}\Psi_{2}+h.c.
\label{eq:eigenE}
\end{eqnarray}
Some studies~\cite{Combescure2009a,Combescure2009b} have shown that block-circulant matrix can be block-diagonalized. So we block-diagonalize the matrix $H_{i}$ to $D_{i}^{\dagger}\widetilde{H}_{i} D_{i}$ and set $\Psi_{X,i}=D_{i}\Psi_{i}$. We have $\widetilde{H}_{i}=diag(\Gamma_{1}^{i},\Gamma_{2}^{i},...,\Gamma_{N_{i}}^{i})$ and $\Psi_{X,1}=(C^{i}_{x,1},C^{i}_{x,2},...,C^{i}_{x,N_{i}})$. Here:
\begin{eqnarray}
\Gamma_{l}^{i}=\sum_{p=1,2,..,N_{i}}exp\left [ -\frac{2i(p-1)(l-1)\pi}{N_i} \right ]a_{p}
\label{eq:Gamma}
\end{eqnarray}
\begin{eqnarray}
C^{i}_{x,l}=\frac{1}{\sqrt{N}}\sum_{p=1,2,..,N}exp\left [ \frac{2i(p-1)(l-1)\pi}{N_i} \right ]C^{i}_{p}
\label{eq:cx}
\end{eqnarray}
As mentioned above, the reordered one dimensional closed cyclic periodic geometry of unit-cells within the supercell supports Bloch-like eigenmodes defined over the finite unit-cell ensemble of each layer. The discrete wave indices $l$ arising from the finite number of unit cells correspond to distinct wave-number sectors, each carrying a different eigenenergy and constituting a discrete band-like energy spectrum. 
A key advantage of our block-diagonalization scheme is that it targets only the inter-unit-cell coupling within the supercell without diagonalizing the intricate Hamiltonian internal to individual primitive cells. This design exclusively addresses the essential supercell formation physics that dominates Moir\'e properties, while bypassing trivial and complicated intra-cell structural details that obscure core Moir\'e-scale phenomena.

In the original atomic basis, interlayer hopping corresponds to direct atomic-scale transitions between the two layers. After the basis transformation, such interlayer couplings are mapped into transition matrix elements between different wave-number sectors residing on the two constituent layers. 
Now we can rewrite Eq.~(~\ref{eq:eigenE}) as $E=\sum_{i}\Psi_{X,i}^{\dagger}\widetilde{H}_{i}\Psi_{X,i}+\Psi_{X,1}^{\dagger}\widetilde{T}_{12}\Psi_{X,2}+h.c.$. The matrix element of $T_{12}$ is $\omega_{s_{1},s_{2}}=[T_{12}]_{s_{1},s_{2}}$ $(s_i=1,2,...,N_i)$. And the matrix element of $\widetilde{T}_{12}$ is $\Omega_{x,y}=[\widetilde{T}_{12}]_{x,y}$. The relationship between $T_{12}$ and $\widetilde{T}_{12}$ is given by:
\begin{equation}
\begin{aligned}
\Omega_{xy}={}&\sum_{s_{i}=1,2,...,N_{i}}\frac{1}{\sqrt{N_{1}N_{2}}}\omega_{s_{1}s_{2}}\\
&\times\exp\left[-2i\pi\left(\frac{(s_{1}-1)(x-1)}{N_{1}}\right.\right.\\
&\quad\left.\left.-\frac{(s_{2}-1)(y-1)}{N_{2}}\right)\right]
\end{aligned}
\label{eq:omega}
\end{equation}
Details of the calculation are provided in the Supplemental Material, Sec. S1.
In this transformed eigenbasis, the original atomic-scale interlayer hopping terms are redistributed across pairwise wave-number sectors between the two layers. Such redistribution fundamentally originates from the ubiquitous lattice mismatch in artificial composite systems, induced by interlayer twist, lattice strain, or dissimilar native lattice constants of the two constituent lattices. Because of this structural incommensurability, interlayer atomic distances and relative stacking positions are spatially irregular rather than uniformly periodic across the supercell. Since interlayer hopping is only significant for atoms with short vertical and in-plane separations, the irregular spatial distribution of atomic overlaps leads to spatially disordered hopping amplitudes throughout the supercell. When projected onto the orderly Bloch wave bases of the two individual layers, these disordered local hopping contributions are broadly scattered and averaged over numerous distinct inter-sector channels. As a result, the effective coupling matrix elements between any pair of specific wave-number sectors become generally small in magnitude. This naturally weakens the interlayer coupling strength in the diagonalized basis and validates the treatment of interlayer hopping as a perturbative correction. 
In quasi-degenerate perturbation theory, only sectors with nearly degenerate energies produce non-negligible perturbative modifications, which allows the original high-dimensional multi-sector lattice problem to be rigorously reduced to a low-dimensional subspace problem dominated by pairwise or several quasi-degenerate wave-number sectors from different layers, greatly simplifying the construction of the effective band Hamiltonian.
To first order in the energy correction, in the case that two energy sectors with different wave-numbers from each layer are degenerated, the Hamiltonian near the degeneracy points reads:
\begin{eqnarray}
H^{d}_{xy}=\begin{bmatrix}
\Gamma^{1}_{x} & \Omega_{xy} \\
\Omega_{xy}^{\dagger} & \Gamma^{2}_{y}
\end{bmatrix}
\label{eq:hd}
\end{eqnarray}
When the zone under consideration contains numerous degeneracy points provided by multiple energy bands, this expression reduces to a continuum model. Throughout our derivation, we do not invoke any low-energy approximation, which means that this continuum model offers the possibility of exploring the high-energy sector as well. Eq.~(\ref{eq:omega}) also does not require $H_{1} \oplus H_{2}$ to be periodic (that is, the moir\'e superlattice structure may, in principle, be quasicrystalline), but if we choose to investigate the band structure in k-space, we are compelled to assume that the moir\'e superlattice is periodic; otherwise, a k-space band description would not be well-defined. 

\subsection{Gauge invariance of the effective Hamiltonian}

The block diagonalization $H_{i}=D_{i}^{\dagger}\widetilde{H}_{i}D_{i}$ is not unique: any $D_{i}$ that diagonalizes $H_{i}$ may be multiplied on the left by a diagonal unitary matrix
\begin{equation}
  U_{i}=\mathrm{diag}(e^{i\phi^{i}_{1}},e^{i\phi^{i}_{2}},\dots,e^{i\phi^{i}_{N_{i}}}),
  \label{eq:gaugeU}
\end{equation}
i.e.\ $D_{i}\to U_{i}D_{i}$, leaving $\widetilde{H}_{i}$ unchanged. Under this freedom the inter-layer coupling matrix $\widetilde{T}_{12}=D_{1}^{\dagger}T_{12}D_{2}$ transforms as
\begin{equation}
  \Omega_{xy}\;\longrightarrow\; e^{-i\phi^{1}_{x}}\,\Omega_{xy}\,e^{i\phi^{2}_{y}},
  \label{eq:gaugeOmega}
\end{equation}
so the individual matrix elements $\Omega_{xy}$ are gauge dependent. The effective Hamiltonian of Eq.~(\ref{eq:hd}), however, transforms as a similarity transformation on the degeneracy subspace,
\begin{equation}
  H^{d}\;\longrightarrow\;
  \begin{bmatrix} U_{1}^{\dagger} & 0 \\ 0 & U_{2}^{\dagger} \end{bmatrix}
  H^{d}
  \begin{bmatrix} U_{1} & 0 \\ 0 & U_{2} \end{bmatrix},
  \label{eq:gaugeHd}
\end{equation}
with $U_{i}$ restricted to the selected degeneracy bands. Consequently the eigenvalue spectrum of $H^{d}$, and every spectral quantity derived from it (density of states, Chern numbers, Berry curvature, and the resulting band topology), is strictly gauge invariant, even though the off-diagonal block $\Omega_{xy}$ itself is not. For practical computations one may fix the gauge by demanding that each degeneracy-point eigenvector of $H_{i}$ be real at some reference position (a real gauge), or by a Wannier-like convention; the conclusions below are independent of this choice.

\subsection{Degenerate Points with Multi-Band Coupling}

Our previous discussion was established under the assumption of two-band degeneracy. Nevertheless, multi-band degeneracy can occur in many systems, such as at the $\Gamma$ point of the TBG system. Now we suppose that the energy bands $\Gamma^{1}_{i_{1}}, \Gamma^{1}_{i_{2}}, ... ,\Gamma^{1}_{i_{m}} $ and $\Gamma^{2}_{j_{1}}, \Gamma^{2}_{j_{2}}, ... ,\Gamma^{2}_{j_{n}} $ are degenerate at the same point. Thus we need to rewrite Eq.~(\ref{eq:hd}) as:
\begin{eqnarray}
H^{d}=\begin{bmatrix}
\Gamma^{1} & \Omega \\
\Omega^{\dagger} & \Gamma^{2}
\end{bmatrix}
\label{eq:hdm}
\end{eqnarray}
Here $\Gamma^{i}=\mathrm{diag}(\Gamma^{1}_{i_{1}}, \Gamma^{1}_{i_{2}}, ... ,\Gamma^{1}_{i_{m}} )$ and $[\Omega]_{ii,jj}=\Omega_{ii,jj}$ ($ii=i_{1},...,i_{n}$, $jj=j_{1},...,j_{m}$).

\subsection{Multilayer Moir\'e pattern}

For an L-layer Moir\'e structure, the Hamiltonian is given by $H^{M}=\sum_{i=1,2,..,L}H_{i}+\sum_{i=1,2,..,L;j=1,2,...,L}T_{ij}$. The eigenvector is $\Psi^{M}=(\Psi_{1}, \Psi_{2}, ..., \Psi_{L})$. Thus, we can rewrite Eq.~(\ref{eq:eigenE}) as
\begin{eqnarray}
E=\sum_{i}\Psi_{i}^{\dagger}H_{i}\Psi_{i}+\sum_{ij}\Psi_{i}^{\dagger}T_{ij}\Psi_{j}+h.c.
\label{eq:eigenEL}
\end{eqnarray}
Our previous calculations show that none of the phases contains the index $i$ of $H_{i}$, therefore, extending the index $i$ from $1,2$ to $1,2,.., L$ does not affect the calculated phases. Thus, we have $\widetilde{H}_{i}=\mathrm{diag}(\Gamma_{1}^{i},\Gamma_{2}^{i},...,\Gamma_{N_{i}}^{i})$ and the interlayer interaction term is:
\begin{equation}
\begin{aligned}
\Omega_{xy}^{ij}={}&\sum_{s_{i},s_{j}}\frac{1}{\sqrt{N_{i}N_{j}}}\omega_{s_{i}s_{j}}^{ij}\\
&\times\exp\left[-2i\pi\left(\frac{(s_{i}-1)(x-1)}{N_{i}}\right.\right.\\
&\quad\left.\left.-\frac{(s_{j}-1)(y-1)}{N_{j}}\right)\right]
\end{aligned}
\label{eq:omegaL}
\end{equation}
These indices also do not affect the first-order perturbation calculation, so the basic results of Eq.~(\ref{eq:hd}) remain unchanged; one simply needs to replace 1 and 2 with the extended indices:
\begin{eqnarray}
H^{d}_{xy}=\begin{bmatrix}
\Gamma^{i}_{x} & \Omega_{xy}^{ij} \\
\Omega_{xy}^{ij\dagger} & \Gamma^{j}_{y}
\end{bmatrix}
\label{eq:hdL}
\end{eqnarray}
An example in case is the twisted bilayer graphene which is
discussed in the following section.

\section{1D-Chain Moir\'e pattern}

In this section, we deal with  speical examples of Mor\'e pattern
formed by two two one-dimensional (1D) atomic chains.

\subsection{Two 1D Atomic Chain With Comensurate Lattice Constants} 

To demonstrate that the construction does not rely on any low-energy approximation, we first apply it to a moir\'e pattern formed by two one-dimensional atomic chains with primitive periods $r_{\alpha}=1$ and $r_{\beta}=4/3$, so that the supercell has period $R=4$ and contains four atoms of chain $\alpha$ and three of chain $\beta$. In this geometry a large number of band degeneracies occur at the same energy, and the inter-chain coupling ($t_{\alpha,\beta}$) strongly modifies the spectrum precisely at those degeneracies while leaving non-degenerate bands essentially unchanged---the hallmark of the perturbative structure of Section~2.

\begin{figure*}[t]
\centering
\includegraphics[width=\textwidth]{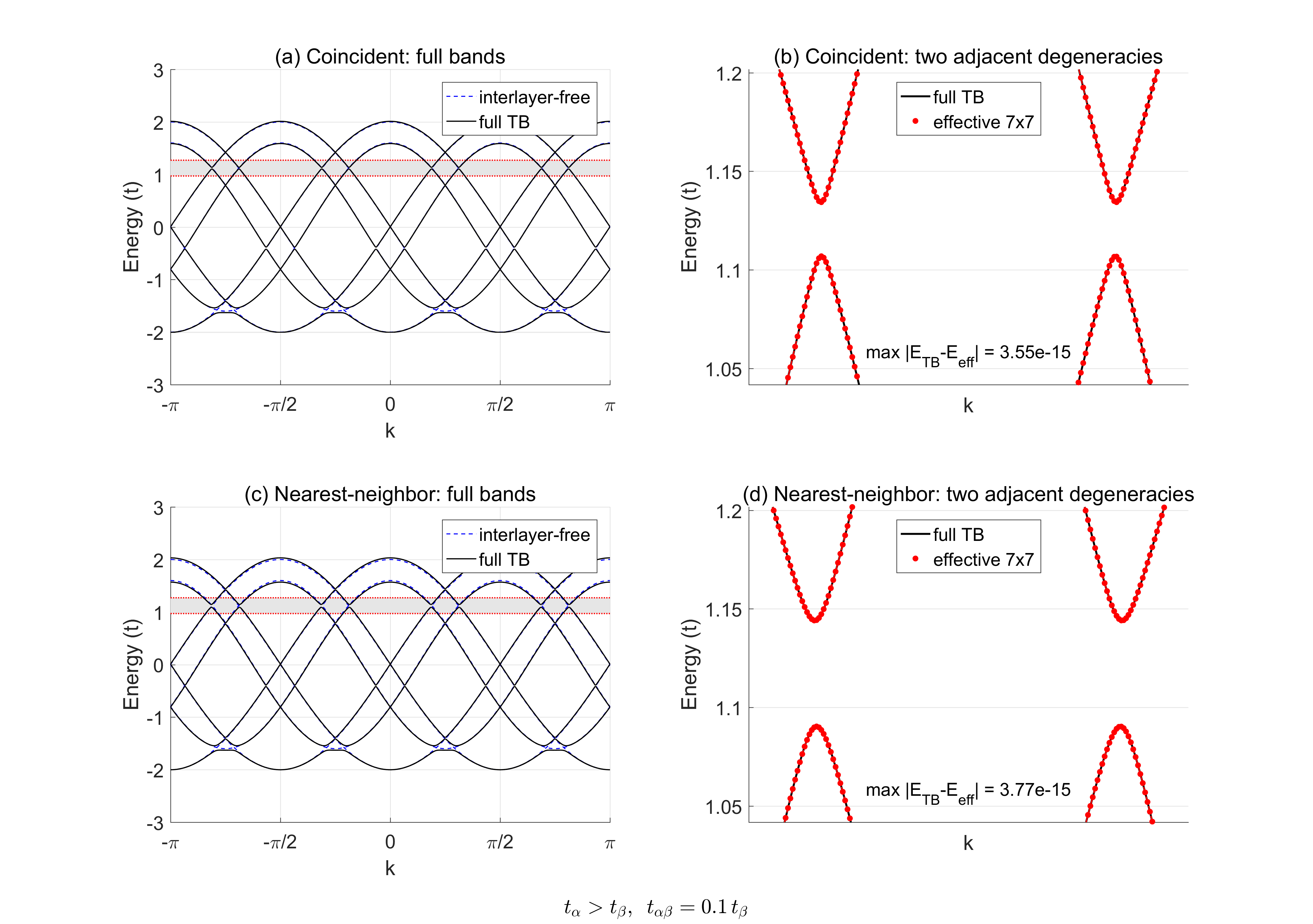}
\caption{Band structure of the two-chain model over the full $k$-range $[-\pi,\pi]$, computed for the two interchain coupling schemes of Sec.~3. (a), (c) Coincident and nearest-neighbor coupling, respectively: the interlayer-free dispersions $\Gamma_{\alpha,i},\Gamma_{\beta,j}$ (blue dashed) together with the full tight-binding bands (black solid); the gray shaded band marks the energy window $[0.97,1.27]$ enclosing the eight degeneracies at $E\approx1.12$. (b), (d) Zoom around two adjacent degeneracy points inside that window: full tight-binding (black solid) versus the $7\times7$ effective model of Eq.~(\ref{eq:1dchain}) (red dashed line with dots). In both schemes the effective model reproduces the full tight-binding spectrum almost exactly. The hopping parameters obey $t_\alpha>t_\beta$ and $t_{\alpha\beta}=0.1\,t_\beta$.}
\label{fig:1dchain}

\end{figure*}

Diagonalizing each intra-chain block by the discrete Fourier transform of the Supplemental Material, Sec.~S1 and keeping only the inter-chain hopping between atoms at coincident positions, the effective Hamiltonian restricted to one degenerate energy window takes the form
\begin{eqnarray}
H(\mathbf{k})=\begin{bmatrix}
\Gamma_{\alpha,1} & \Omega & \Omega & 0 & 0 & 0 & 0 \\
\Omega & \Gamma_{\beta_2} & 0 & \Omega & \Omega & 0 & 0 \\
\Omega & 0 & \Gamma_{\beta,3} & \Omega & 0 & \Omega & 0 \\
0 & \Omega & \Omega & \Gamma_{\alpha_{3}} & 0 & 0 & 0 \\
0 & \Omega & 0 & 0 & \Gamma_{\alpha_{4}} & 0 & \Omega \\
0 & 0 & \Omega & 0 & 0 & \Gamma_{\alpha_{2}} & \Omega \\
0 & 0 & 0 & 0 & \Omega & \Omega & \Gamma_{\beta_{1}} \\
\end{bmatrix}
\label{eq:1dchain}
\end{eqnarray}
with diagonal entries $\Gamma_{\alpha,i}=2t_{\alpha}\cos(\mathbf{k}\cdot\mathbf{r_{\alpha}}+2(i-1)\pi/4)$, $\Gamma_{\beta,i}=2t_{\beta}\cos(\mathbf{k}\cdot\mathbf{r_{\beta}}+2(i-1)/3)$, and $\Omega=t_{\alpha,\beta}/\sqrt{12}$; the full derivation is given in the Supplemental Material, Sec.~S2. This Hamiltonian is \emph{finite}, not continuum, because it retains only the degeneracies inside the selected energy window (the shaded band of Fig.~\ref{fig:1dchain}(a)), which is not a low-energy zone. Figs.~\ref{fig:1dchain}(b) and (c) show that, within that window and around a single degeneracy point, the effective Hamiltonian reproduces the full tight-binding band structure almost exactly.

The effective Hamiltonian of Eq.~(\ref{eq:1dchain}) retains only the inter-chain hopping between atoms at coincident in-plane positions. A more faithful microscopic picture lets every atom couple to the \emph{nearest} atom of the other chain, with the coupling amplitude fixed by the in-plane offset; we outline this generalization here. With the chain atoms at positions $x_{\alpha}(s)=(s-1)r_{\alpha}$ ($s=1,\dots,4$) and $x_{\beta}(t)=(t-1)r_{\beta}$ ($t=1,2,3$), the nearest-neighbor pairs inside one supercell (modulo the period $R=4$) are
\begin{equation}
(s,t)\;\in\;\big\{(1,1),\ (2,2),\ (3,2),\ (3,3),\ (4,3)\big\},
\label{eq:nnpairs}
\end{equation}
with in-plane offsets $\Delta_{st}=x_{\beta}(t)-x_{\alpha}(s)$ equal to $0$, $\tfrac13$, $-\tfrac23$, $\tfrac23$, $-\tfrac13$, respectively; note that $\alpha_3$ possesses two equidistant nearest neighbors, $\beta_2$ and $\beta_3$. The inter-chain coupling in the supercell Bloch basis is then
\begin{equation}
H_{\mathrm{int}}=\sum_{(s,t)} t_{\alpha\beta}\big(c_{\alpha,s}^{\dagger}c_{\beta,t}\,e^{ik\Delta_{st}}+\mathrm{h.c.}\big),
\end{equation}
where the phase $e^{ik\Delta_{st}}$ accounts for the in-plane offset between the two coupled atoms. Folding this through Eq.~(\ref{eq:omega}) with $\omega_{st}=t_{\alpha\beta}e^{ik\Delta_{st}}$ gives the momentum-dependent inter-chain block
\begin{multline}
\Omega_{ij}(k)=\frac{1}{\sqrt{N_{\alpha}N_{\beta}}}\sum_{(s,t)} t_{\alpha\beta}\,e^{ik\Delta_{st}}\\
\times e^{-2\pi i\left[\frac{(s-1)(i-1)}{N_{\alpha}}-\frac{(t-1)(j-1)}{N_{\beta}}\right]},
\label{eq:omegaNN}
\end{multline}
which, together with the diagonal blocks $\Gamma_{\alpha,i}(k)$ and $\Gamma_{\beta,j}(k)$ of Eq.~(\ref{eq:1dchain}), assembles the full $7\times7$ effective Hamiltonian. In contrast to the coincident case, where the folded block reduces to the constant $\Omega=t_{\alpha\beta}/\sqrt{12}$, the nearest-neighbor block of Eq.~(\ref{eq:omegaNN}) is explicitly $k$-dependent. Figs.~\ref{fig:1dchain}(d)--(f) show the corresponding bands for the nearest-neighbor coupling; in both schemes the $7\times7$ effective Hamiltonian reproduces the full tight-binding spectrum over the entire $k$-range.

\subsection{Two 1D Atomic Chain With Imcommensurate Lattice Constants}

We consider a bilayer one‑dimensional moir\'e system composed of two atomic chains. The first chain $\alpha$ possesses lattice constant $a_1=1$, with lattice sites located at $x_\alpha(n)=n$, $n=1,2,\dots,N_1$ and nearest‑neighbor intralayer hopping amplitude $t_\alpha$. The second chain $\beta$ has lattice constant $a_2=\alpha_0$, where $\alpha_0$ is an irrational number, with lattice sites $x_\beta(m)=m\alpha_0$, $m=1,2,\dots,N_2$ and intralayer hopping $t_\beta$. Interlayer hopping $t_\perp$ occurs only between spatially proximal sites satisfying $|x_\alpha(n)-x_\beta(m)|<\delta$ (usually $\delta$ can be chosen to one or two lattice constants), and vanishes otherwise. Since $\alpha_0$ is irrational, no finite commensurate supercell exists for the genuine incommensurate system. We adopt a sequence of commensurate approximants constructed from continued‑fraction convergents $p_k/q_k$ of $\alpha_0$, such that $\lim_{k\to\infty} p_k/q_k=\alpha_0$. For each finite approximant index $k$, we set supercell sizes $N_1=p_k$, $N_2=q_k$, which defines a periodic surrogate system approximating the target incommensurate bilayer.

\begin{figure}[htbp]
\centering
\begin{tikzpicture}[
    atom/.style={circle,fill,minimum size=4pt,inner sep=0pt},
    intralayer/.style={thick},
    interlayer/.style={dashed,thick,blue!70},
    >=Stealth
]

\foreach \n in {0,1,2,3,4,5}{
    \node[atom,red] (a\n) at (\n, 1.2) {};
}
\foreach \n in {0,...,4}{
    \draw[intralayer] (a\n) -- (a\the\numexpr\n+1);
}
\node[above=3pt of a2] {Chain $\alpha$};
\node[above=6pt of a0] {$x_\alpha(n)=n a_1$};
\draw[<->] (a0) -- (a1) node[midway,above=2pt] {$a_1$};
\node[red,above=6pt of a4] {$t_\alpha$};

\def\ax{1.414}
\foreach \m in {0,1,2,3}{
    \pgfmathsetmacro{\xm}{\m*\ax}
    \node[atom,black] (b\m) at (\xm, 0) {};
}
\draw[intralayer] (b0)--(b1)--(b2)--(b3);
\node[below=3pt of b3] {Chain $\beta$};
\node[below=3pt of b0] {$x_\beta(m)=m a_2$};
\draw[<->] (b1) -- (b2) node[midway,below=2pt] {$a_2$};
\node[black,below=6pt of b2] {$t_\beta$};

\draw[interlayer] (a0) -- (b0) node[midway,right=2pt,blue] {$t_\perp$};
\draw[interlayer] (a1) -- (b0);
\draw[interlayer] (a1) -- (b1);
\draw[interlayer] (a2) -- (b1);
\draw[interlayer] (a2) -- (b2);
\draw[interlayer] (a3) -- (b2);
\draw[interlayer] (a4) -- (b3);
\draw[interlayer] (a5) -- (b3);

\draw[<->,green!80] 
    node[midway,left=4pt,green] {$\delta$};

\end{tikzpicture}
\caption{Schematic diagram of the bilayer one‑dimensional moiré lattice. Two atomic chains $\alpha$ (red) and $\beta$ (black) possess lattice constants $a_1$ and $a_2=\alpha_0$, where $\alpha_0$ is irrational. $t_\alpha$ and $t_\beta$ denote intralayer nearest‑neighbor hopping amplitudes within each individual chain. Interlayer hopping $t_\perp$ (blue dashed lines) is present only for spatially proximal atomic pairs satisfying $|x_\alpha(n)-x_\beta(m)|<\delta$, and vanishes otherwise. Since $a_2/a_1$ is irrational, this system is strictly incommensurate and possesses no finite supercell.}
\label{fig:bilayer_1d_moire}
\end{figure}
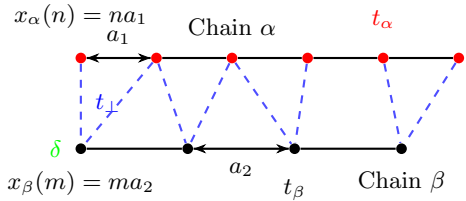

For each approximant, we employ block‑circulant diagonalization via discrete Fourier transform to transform the intralayer Hamiltonians into their Bloch eigenbasis. The real‑space interlayer hopping matrix elements are folded into this Bloch basis, yielding the off‑diagonal interlayer coupling matrix
\begin{equation}
\Omega_{xy}^{(k)}=\frac{t_\perp}{\sqrt{N_1 N_2}}\sum_{n=1}^{N_1}W_k(n)\exp\left\{-i\left(k_x^{(k)}-\frac{k_y^{(k)}}{\alpha_0}\right)n\right\},
\label{eq:omega_xy_k}
\end{equation}
where $k_x^{(k)}=2\pi(x-1)/N_1$, $k_y^{(k)}=2\pi(y-1)/N_2$ are discrete wave‑numbers for chain‑$\alpha$ and chain‑$\beta$, respectively, and the real‑space matching‑weight function
$W_k(n)=1$ if there exists $m$ satisfying $|n-m\alpha_0|<\delta$, and zero otherwise. 

In the high‑order limit $k\to\infty$, $N_1,N_2\to\infty$. Owing to Weyl's equidistribution theorem, the set of site indices satisfying the spatial proximity condition becomes uniformly distributed, and the weight function $W_k(n)$ converges in the distributional sense to a quasi‑periodic function dominated by the moir\'e wave‑vector $q=2\pi(1-1/\alpha_0)$, i.e., $W_k(x)\xrightarrow{\text{dist}} C\cos(qx)$ (See supplementary material Sec.(\ref{app:C}) for detailed calculations). The artificial periodicity inherited from the finite‑size supercell of each commensurate approximant is lifted in this limit. Substituting this asymptotic form for $W_k$, the projected folded coupling $[\Omega_{\mathcal S_k}]_{xy}$ converges under the strong‑operator topology to the multiplication‑type quasi‑periodic potential $V(x)=2t_\perp\cos(qx)$, where $q/(2\pi)$ remains irrational, guaranteeing genuine incommensurability. Performing long‑wave expansion for the diagonal tight‑binding dispersions and absorbing constant energy offsets, the whole finite‑dimensional effective Hamiltonian $H_k$ constructed from block‑circulant diagonalization and quasi‑degenerate projection converges to the continuous matrix‑valued Almost‑Mathieu operator
\begin{equation}
H_{\rm MAM}=
\begin{pmatrix}
-t_\alpha\partial_x^2 & 2t_\perp\cos(qx)\\
2t_\perp\cos(qx) & -t_\beta\partial_x^2
\end{pmatrix}.
\label{eq:MAM_operator}
\end{equation}

This operator describes two coupled one‑dimensional channels subject to an incommensurate quasi‑periodic inter‑channel coupling.
Since $\alpha_0$ is irrational, the moiré wavevector $q$ is incommensurate with the lattice wavevector, yielding a quasiperiodic potential without any finite supercell. This incommensurate modulation opens infinitely many energy gaps across all scales, repeatedly fragmenting the bands and producing a self-similar fractal Cantor spectrum. This case happens in the strong‑interlayer‑coupling regime above the self‑dual critical coupling $t_\perp>(t_\perp)_c$, the spectrum of $H_{\rm MAM}$ forms a fractal Cantor set with zero Lebesgue measure, which can be obtained as the topological closure over the spectra from the full sequence of commensurate approximant Hamiltonians $\{H_k\}$. Corresponding to this Cantor spectrum, the Lyapunov exponents computed via transfer‑matrix techniques are strictly positive across the full energy spectrum, signalling complete exponential localization of all eigenstates. At the critical point $t_\perp=(t_\perp)_c$, the system exhibits singular‑continuous spectra with intermediate fractal dimension, while for weak coupling  $t_\perp<(t_\perp)_c$, extended states appear and the spectrum possesses continuous spectral components. These results are typical
for imcomensurate lattice and what we want to emphasize here is
the Eq.(\ref{eq:omega_xy_k}) which is a consequence of our general
analytic model applied to imcommensurate lattice.

\begin{figure*}[t]
\centering
\includegraphics[width=\textwidth]{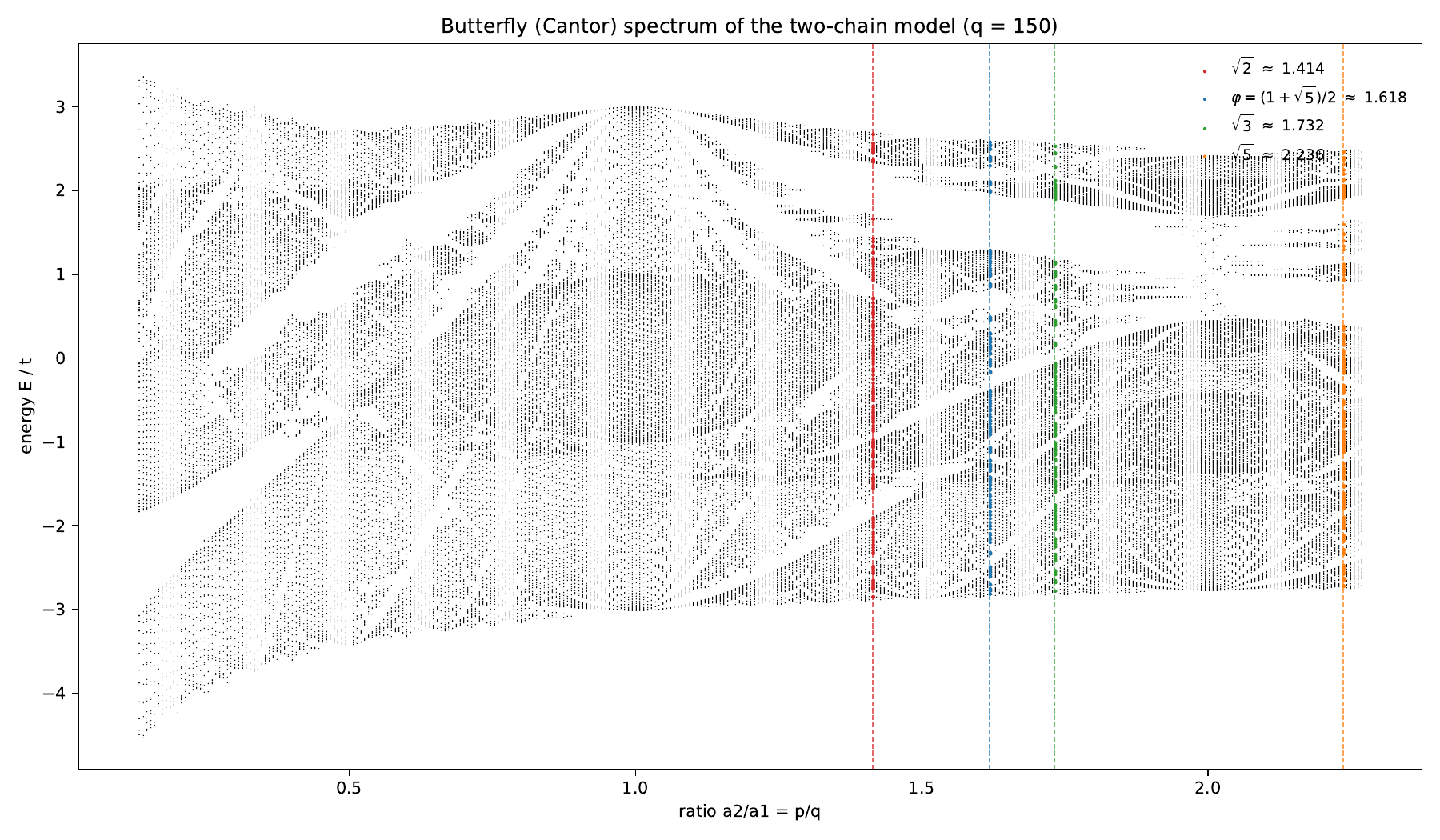}
\caption{Butterfly (Cantor) spectrum of the two-chain model, obtained by
diagonalizing the $(p+q)\times(p+q)$ commensurate-approximant Hamiltonian as
the rational ratio $p/q$ is swept at fixed $q=150$. Four badly approximable
quadratic irrationals are marked with thicker colored dots---$\sqrt2$ (red),
$\varphi=(1+\sqrt5)/2$ (blue), $\sqrt3$ (green), and $\sqrt5$ (orange). The
box-counting dimension of Fig.~\ref{fig:box_counting} is evaluated at
$\varphi$.}
\label{fig:cantor_butterfly}
\end{figure*}

The fractal character of this spectrum is readily confirmed numerically: the
band fragmentation is already apparent from a direct diagonalization of the
commensurate-approximant two-chain Hamiltonian. We choose the golden mean
$\varphi=(1+\sqrt5)/2$ as the irrational ratio, whose rational approximants
converge more slowly than those of any other number, so that the self-similar
splitting is maximally regular. Figure~\ref{fig:cantor_butterfly} shows the
eigenvalue spectrum of the $(p+q)\times(p+q)$ Hamiltonian as the rational
ratio $p/q$ is swept at fixed $q=150$; the resulting Hofstadter-like butterfly
develops gaps on every scale. Four quadratic irrationals---$\sqrt2$,
$\varphi$, $\sqrt3$, and $\sqrt5$---are highlighted, each being badly
approximable and each seeding a dense Cantor-like fragmentation that is
sharpest at $\varphi$.

\begin{figure}[htbp]
\centering
\includegraphics[width=\columnwidth]{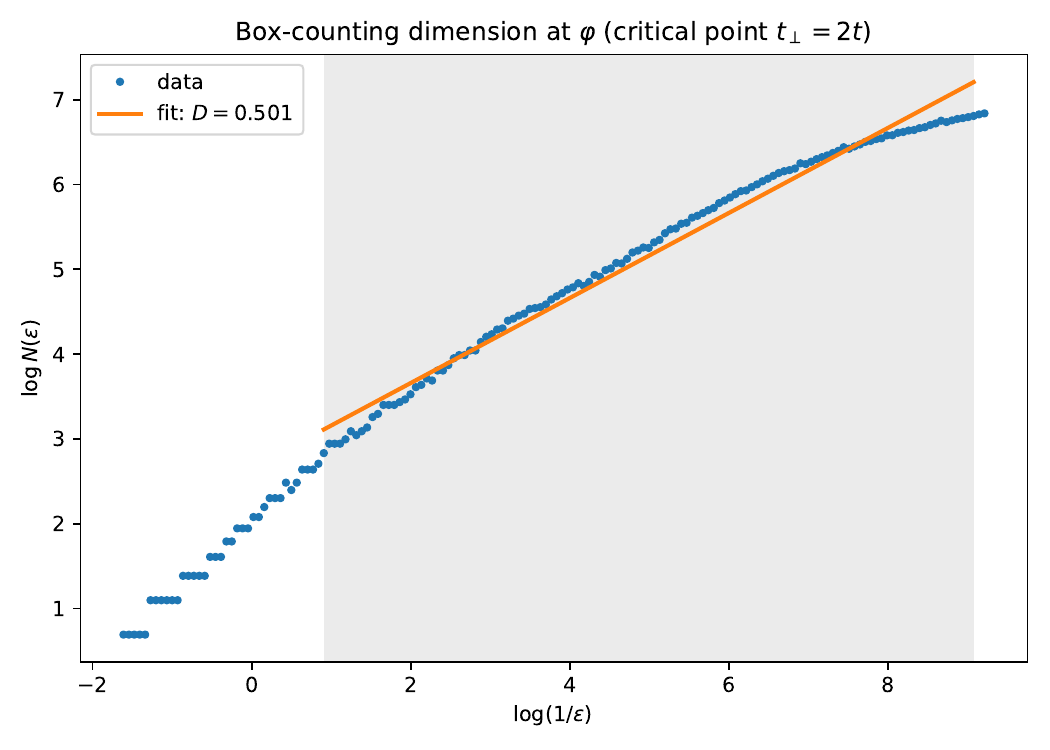}
\caption{Box-counting dimension of the two-chain spectrum at the golden ratio
$\varphi$, evaluated at the critical coupling $t_\perp=2t$ using the Fibonacci
approximant $1597/987$. The shaded band marks the fractal window located
automatically from the intermediate running slope; the linear fit (red) gives
$D\approx0.5$.}
\label{fig:box_counting}
\end{figure}

To assign a quantitative measure to this fragmentation we evaluate the
box-counting (Minkowski) dimension at the self-dual point. Approximating
$\varphi$ by the Fibonacci ratio $1597/987$ and setting $t_\perp=2t$, we count
the number $N(\varepsilon)$ of bins of width $\varepsilon$ required to cover
the spectrum, for which $N(\varepsilon)\propto\varepsilon^{-D}$. The slope of
$\log N$ against $\log(1/\varepsilon)$ is trustworthy only in an intermediate
range of $\varepsilon$: bins much wider than the largest gaps do not resolve
the Cantor structure (the slope tends to unity), whereas bins much narrower
than the mean level spacing contain at most one eigenvalue (the slope tends to
zero). We therefore locate the fractal window by tracking the running
logarithmic derivative, smoothing it, and keeping the longest interval over
which the smoothed derivative lies between $0.15$ and $0.8$; a least-squares
fit within this window [Fig.~\ref{fig:box_counting}] yields $D\approx0.5$,
i.e.\ the H\"older exponent $1/2$ of the critical integrated density of
states, consistent with the zero-measure Cantor structure of the critical
spectrum.

\section{1D Atomic Chain Embedded in a 2D Square Lattice}
In this section, we consider a case in which a 1D atomic chain
is embedded on top of a 2D square lattce.

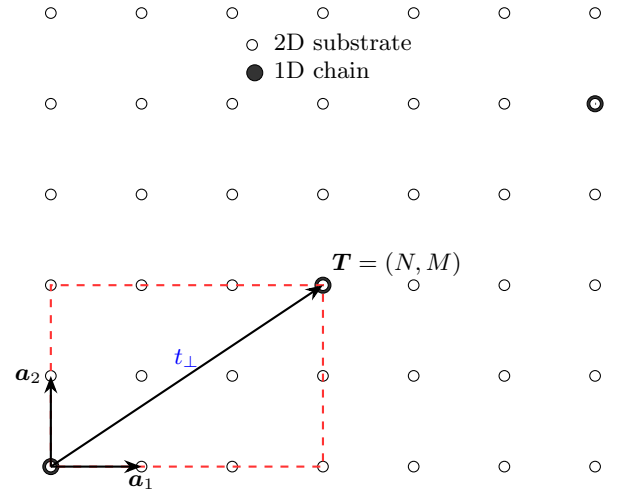
\begin{figure}[htbp]
\centering
\begin{tikzpicture}[
    >=Stealth,
    substrate/.style={circle,draw,fill=white,minimum size=4pt,inner sep=0pt},
    chainatom/.style={circle,draw,fill=black!80,minimum size=6pt,inner sep=0pt},
    dashcouple/.style={black!70,dashed,thick}
]
\def\N{3}
\def\M{2}
\def\a{1.2}
\def\shiftx{\a}
\def\shifty{\a}

\foreach \x in {0,...,6}{
    \foreach \y in {0,...,5}{
        \pgfmathsetmacro{\xx}{\x*\a + \shiftx}
        \pgfmathsetmacro{\yy}{\y*\a + \shifty}
        \node[substrate] at (\xx,\yy) {};
    }
}

\def\nmax{2}
\foreach \n in {0,...,\nmax}{
    \pgfmathsetmacro{\cx}{\n*\N*\a + \shiftx}
    \pgfmathsetmacro{\cy}{\n*\M*\a + \shifty}
    \node[chainatom] (c\n) at (\cx,\cy) {};
}

\foreach \n in {0,...,\nmax}{
    \pgfmathsetmacro{\sx}{\n*\N*\a + \shiftx}
    \pgfmathsetmacro{\sy}{\n*\M*\a + \shifty}
    \node[substrate] (s\n) at (\sx,\sy) {};
    \draw[dashcouple] (c\n.center) -- (s\n.center);
}

\pgfmathsetmacro{\boxx}{\N*\a+\shiftx}
\pgfmathsetmacro{\boxy}{\M*\a+\shifty}
\draw[red!80,dashed,thick] (\shiftx,\shifty) rectangle (\boxx,\boxy);

\draw[->,thick] (\shiftx,\shifty) -- (\a+\shiftx,\shifty) node[below] {$\boldsymbol a_1$};
\draw[->,thick] (\shiftx,\shifty) -- (\shiftx,\a+\shifty) node[left] {$\boldsymbol a_2$};
\draw[->,thick] (\shiftx,\shifty) -- (\boxx,\boxy) node[above right] {$\boldsymbol T=(N,M)$};


\node[blue,above] at ({0.5*\N*\a+\shiftx},{0.5*\M*\a+\shifty}) {$t_\perp$};

\node[anchor=west] at ({2.0*\a+\shiftx},{4.5*\a+\shifty}) {
\begin{tabular}{l l}
\tikz[baseline=-0.5ex]{\node[substrate]{};} & 2D substrate \\
\tikz[baseline=-0.5ex]{\node[chainatom]{};} & 1D chain
\end{tabular}
};
\end{tikzpicture}

\caption{Schematic of a one-dimensional atomic chain embedded on a two-dimensional square lattice.
$\boldsymbol a_1,\boldsymbol a_2$ are primitive lattice vectors of the substrate.
$\boldsymbol T=(N,M)$ denotes the translation vector of the 1D chain.
$t_\perp$ represents the interlayer hopping between the chain and substrate.}
\label{fig:1d_chain_2d_sub}
\end{figure}

We consider a 2D square lattice with lattice constant $a=1$, primitive vectors $\boldsymbol{a}_1=(1,0)$ and $\boldsymbol{a}_2=(0,1)$, and reciprocal vectors $\boldsymbol{b}_1=(2\pi,0)$, $\boldsymbol{b}_2=(0,2\pi)$. The substrate lattice sites read $\boldsymbol{R}_{m_1m_2}=(m_1,m_2)$ with integer indices $m_1,m_2$, and we denote substrate electron operators as $c^\dagger_{\boldsymbol{R}},c_{\boldsymbol{R}}$.

A 1D tight-binding chain is straightly embedded in the plane with lattice sites $\boldsymbol{r}_n=\boldsymbol{r}_0+n\boldsymbol{T}$, where $\boldsymbol{T}=(N,M)$ denotes a translation vector with \textit{coprime integers} $N,M$, ensuring a \textit{commensurate configuration}. The chain creation/annihilation operators are $d^\dagger_n,d_n$.

We consider the commensurate case, which means supercell exist in
such system. The minimal commensurate supercell is spanned by the substrate primitive vector $\boldsymbol{a}_1$ and the chain translation $\boldsymbol{T}=(N,M)$. The chain possesses a rational slope $M/N$, exhibiting periodic registration with the 2D lattice after each integer translation $(N,M)$. This periodicity guarantees a finite supercell, validating Bloch’s theorem for the full composite system and yielding a finite-band spectrum.

The total Hamiltonian decomposes into three components $H=H_{2D}+H_{1D}+H_{\text{couple}}$. The 2D substrate tight-binding Hamiltonian with nearest-neighbor hopping $t_2$ is
\[
H_{2D}=-t_2\sum_{\langle\boldsymbol{R},\boldsymbol{R}^\prime\rangle}\big(c^\dagger_{\boldsymbol{R}}c_{\boldsymbol{R}^\prime}+\text{h.c.}\big),
\]
supporting the dispersion $\varepsilon_{2D}(\boldsymbol{k})=-2t_2(\cos k_x+\cos k_y)$ for $\boldsymbol{k}=(k_x,k_y)$. The embedded 1D chain with hopping $t_1$ reads
\[
H_{1D}=-t_1\sum_n\big(d^\dagger_n d_{n+1}+\text{h.c.}\big),
\]
with 1D dispersion $\varepsilon_{1D}(p)=-2t_1\cos p$, where $p$ is the Bloch momentum along the chain.

We adopt a short spatial cutoff $1$, such that each chain site only couples to its coinciding substrate site with interband hopping strength $t_\perp$. Setting zero offset $\boldsymbol{r}_0=\boldsymbol{0}$, we have $\boldsymbol{r}_n=(nN,nM)$, and the coupling Hamiltonian takes the compact form
\[
H_{\text{couple}}=t_\perp\sum_n\big(d^\dagger_n c_{nN,nM}+\text{h.c.}\big).
\]

We Fourier transform the lattice operators as $c_{\boldsymbol{R}}=\int d^2k/(2\pi)^2\,e^{i\boldsymbol{k}\cdot\boldsymbol{R}}c_{\boldsymbol{k}}$ and $d_n=\int dp/(2\pi)\,e^{ipn}d_p$. Substituting into $H_{\text{couple}}$ gives
\[
H_{\text{couple}}=t_\perp\sum_n\int\frac{d^2kdp}{(2\pi)^3}\big(d^\dagger_p c_{\boldsymbol{k}}e^{in(Nk_x+Mk_y-p)}+\text{h.c.}\big).
\]
Applying the Poisson summation formula $\sum_n e^{in\theta}=2\pi\sum_G\delta(\theta+2\pi G)$, we obtain the \textit{resonance selection rule} within the first Brillouin zone ($G=0$): $p=Nk_x+Mk_y$.

The fundamental momentum-space \textit{interband hopping matrix element} is
\begin{equation}
\Omega(p,\boldsymbol{k})=2\pi t_\perp\,\delta\left(p-Nk_x-Mk_y\right),
\end{equation}
which demonstrates that finite chain-substrate transition only occurs on the resonance momentum manifold $p=Nk_x+Mk_y$.

Projecting the full system onto the resonance manifold yields the minimal two-band effective Hamiltonian
\begin{multline}
H_{\mathrm{eff}}(\boldsymbol{k})\\
=\begin{pmatrix}
-2t_1\cos(Nk_x+Mk_y) & t_\perp \\
t_\perp & -2t_2(\cos k_x+\cos k_y)
\end{pmatrix}.
\end{multline}
Analytic diagonalization produces the closed-form band dispersions
\begin{equation}
    E_\pm(\boldsymbol{k})=\frac{\varepsilon_{1D}+\varepsilon_{2D}}{2}\pm\sqrt{\left(\frac{\varepsilon_{1D}-\varepsilon_{2D}}{2}\right)^2+t_\perp^2},
\end{equation}
where $\varepsilon_{1D}=-2t_1\cos(Nk_x+Mk_y)$ and $\varepsilon_{2D}=-2t_2(\cos k_x+\cos k_y)$. A band gap $\Delta=2t_\perp$ is opened at the degeneracy condition $\varepsilon_{1D}=\varepsilon_{2D}$.

For general commensurate embedding with coprime $(N,M)$, the finite supercell guarantees a \textit{finite number of continuous bands and finite gaps}, with analytic band solutions available.

For incommensurate embedding with irrational slope $M/N$, no finite supercell exists. Another case of imcommensurate is the although there exist a rational slope $M/N$, the lattice constant of 1D chain
is an irrational ratio with respect to the period of 2D lattice projected on the 1D chain. In such case,  the low-energy physics reduces to an incommensurate Almost-Mathieu quasiperiodic operator, generating infinite hierarchical avoided crossings, self-similar gap structures, and a fractal Cantor-set spectrum.

\section{2D Moir\'e pattern}

In this section, we investigate moiré patterns formed by two stacked two-dimensional lattices, where one lattice overlaps the other to construct generic bilayer composite lattice structures.
We firstly shown that in what condition, the unit cells in the two-dimensional supercell can be reordered into one-dimensional chain.

\subsection{Transform from 2D to 1D}
\label{sec31}
\begin{figure}[ht]
\centering
\includegraphics[width=\linewidth]{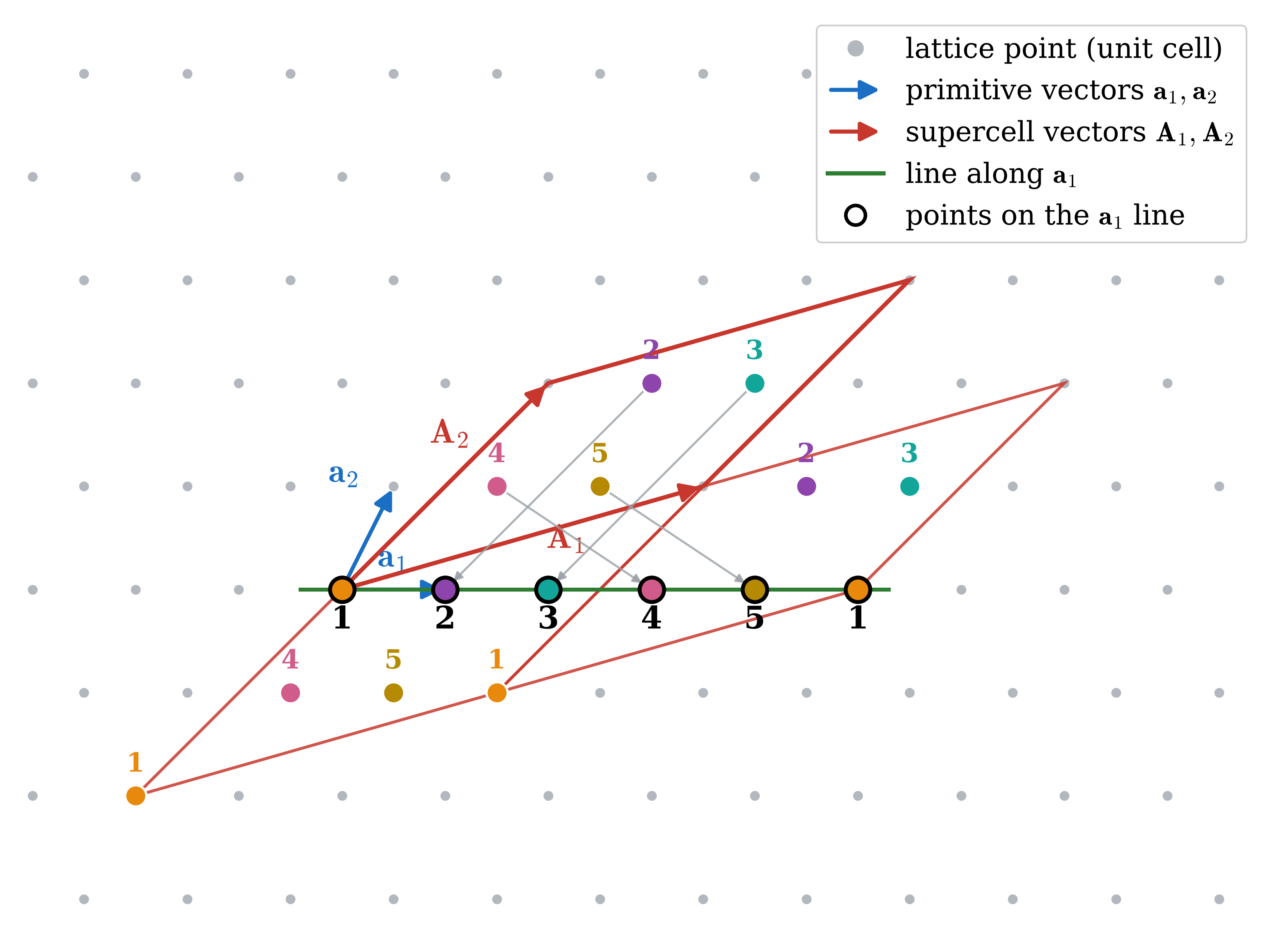}
\caption{Example of the 2D$\to$1D mapping. Each gray point is a primitive cell; the red parallelogram is the supercell. Using the periodicity of $\mathbf{A}_{1},\mathbf{A}_{2}$, every cell in the supercell is identified with a distinct cell on the green line (the generator direction).}
\label{fig:2dto1d}

\end{figure}

Consider a two-dimensional lattice with primitive vectors $\mathbf{a}_{1},\mathbf{a}_{2}$, and a supercell spanned by $\mathbf{A}_{1}=m_{1}\mathbf{a}_{1}+n_{1}\mathbf{a}_{2}$ and $\mathbf{A}_{2}=m_{2}\mathbf{a}_{1}+n_{2}\mathbf{a}_{2}$, containing $N_{c}=|m_{1}n_{2}-m_{2}n_{1}|$ primitive cells. We prove that, under the coprime conditions $\gcd(m_{1},m_{2})=\gcd(n_{1},n_{2})=1$, the $N_{c}$ cells form a single cyclic chain. Let
\begin{equation}
S=\begin{pmatrix}m_{1}&m_{2}\\n_{1}&n_{2}\end{pmatrix}\in\mathbb{Z}^{2\times2},
\qquad N_{c}=|\det S|.
\end{equation}
The distinct cells inside one supercell are the elements of the quotient group $G=\mathbb{Z}^{2}/S\mathbb{Z}^{2}$, whose order is $N_{c}$. By the Smith normal form~\cite{Smith1861}, there exist unimodular $U,V$ such that $USV=\mathrm{diag}(d_{1},d_{2})$ with $d_{1}d_{2}=N_{c}$ and $d_{1}=\gcd(m_{1},n_{1},m_{2},n_{2})=1$. Hence $G\cong\mathbb{Z}_{N_{c}}$ is cyclic, generated by $\mathbf{g}=[\mathbf{v}]$. Numbering the cells as $k\mapsto k\mathbf{g}$ ($k=0,\dots,N_{c}-1$) with representative positions $\mathbf{r}_{k}=k\mathbf{v}+\mathbf{r}_{0}\pmod{\Lambda_{\mathrm{sup}}}$, the cells close cyclically since $N_{c}\mathbf{v}\in\Lambda_{\mathrm{sup}}$. The hopping matrix element $H_{ij}=t(\mathbf{r}_{j}-\mathbf{r}_{i})$ then depends only on $(j-i)\bmod N_{c}$, so $H$ is block-circulant; the two-dimensional supercell is thereby turned into an effective one-dimensional chain (Fig.~\ref{fig:2dto1d}).

To identify the displacement $\mathbf{A}$ and the chain factor $f$, choose the generator vector $\mathbf{A}=n_{2}\mathbf{A}_{1}-n_{1}\mathbf{A}_{2}=(m_{1}n_{2}-m_{2}n_{1})\mathbf{a}_{1}$, which traverses all $N_{c}$ cells. A cell $O_{l}$ at $m\mathbf{a}_{1}+n\mathbf{a}_{2}$ is equivalent (modulo the supercell) to $(m+\alpha m_{1}+\beta m_{2})\mathbf{a}_{1}+(n+\alpha n_{1}+\beta n_{2})\mathbf{a}_{2}$; requiring $n+\alpha n_{1}+\beta n_{2}=0$ determines $(\alpha,\beta)$, and the chain index is $l=(m+\alpha m_{1}+\beta m_{2})\bmod N_{c}$. One then shows that consecutive cells along $\mathbf{A}$ satisfy $\mathbf{O}_{l+1}-\mathbf{O}_{l}=\mathbf{a}_{1}$ and $\mathbf{O}_{l+f}-\mathbf{O}_{l}=\mathbf{a}_{2}$, with $f$ a constant independent of $l$. Applying the discrete Fourier transform of Section~2 to the block-circulant Hamiltonian with at most seven nearest-cell terms $a_{1},a_{2},a_{0}=a_{N_{c}},a_{f},a_{-f},a_{f-1},a_{-(f-1)}$ gives
\begin{equation}
\begin{aligned}
    \Gamma_{l}=a_{1}+e^{-\frac{2\pi i(l-1)}{N_{c}}}a_{2}
+e^{-\frac{2\pi i(f-1)(l-1)}{N_{c}}}a_{f}\\
 \ \   + e^{-\frac{2\pi i(f-2)(l-1)}{N_{c}}}a_{f-1}+h.c.,
 \end{aligned}
\end{equation}
and analogously $\Gamma_{l'}'$ for the second layer; the inter-layer block is
\begin{equation}
\Omega_{ll'}=\sum_{s_{\alpha},s_{\beta}}\frac{1}{\sqrt{N_{c}N_{c}'}}
e^{-2\pi i\left(\frac{(s_{\alpha}-1)(l-1)}{N_{c}}-\frac{(s_{\beta}-1)(l'-1)}{N_{c}'}\right)}\omega_{s_{\alpha}s_{\beta}}.
\end{equation}

\subsection{Special case: twisted bilayer graphene}
\label{sec42}
For twisted bilayer graphene, the two layers share the same supercell, and the commensurate geometry fixes the supercell vectors, the cell number, the chain factor $f$, and the supercell reciprocal vectors $\mathbf{B}_{1},\mathbf{B}_{2}$. These geometric ingredients (derived in the Supplemental Material, Sec.~S3) are:
\begin{equation}
\begin{aligned}
\mathbf{A}_{1}&=-2n\mathbf{a}_{1}+(n+1)\mathbf{a}_{2},\\
\mathbf{A}_{2}&=-(n+1)\mathbf{a}_{1}+(-n+1)\mathbf{a}_{2},
\end{aligned}
\end{equation}
with twist angle $\theta=\sqrt{3}/(3n)$ and $N_{c}=3n^{2}+1$; the chain factor is
\begin{equation}
f=\frac{3}{2}n(n+1)+1,
\end{equation}
and the supercell reciprocal vector $\mathbf{B}_{1}$ is characterized by the modular inverse relation
\begin{equation}
\begin{aligned}
\mathbf{B}_{2}&\equiv c\,\mathbf{B}_{1}\pmod{\mathbf{b}_{1},\mathbf{b}_{2}},\\
c&\equiv -(n+1)\,(-n+1)^{-1}\pmod{N_{c}}.
\end{aligned}
\label{eq:cvalue}
\end{equation}
Crucially, $(-n+1)^{-1}$ is a modular inverse of order $N_{c}$, not the real number $1/n$ (see the remark in the Supplemental Material, Sec.~S3).

Applying the discrete Fourier transform of Section~2 to the inter-layer coupling (the full derivation is in the Supplemental Material, Sec.~S4), we can find out that the DFT phase is a $\mathbf{G}\cdot\mathbf{r}$ dot product, the folded inter-layer block is diagonal in the folded momentum, and momentum conservation selects the three minimal transfers $\mathbf{q}_{1},\mathbf{q}_{2},\mathbf{q}_{3}$ (length $k_{\theta}=2k_{D}\sin(\theta/2)$) connecting the $\alpha$-layer Dirac point to its three nearest $\beta$-layer counterparts. The result is the Bistritzer--MacDonald form
\begin{equation}
\Omega_{ll'}=\sum_{j=1}^{3} T_j\,\delta_{\delta\mathbf{k}_{\alpha}-\delta\mathbf{k}_{\beta},\;\mathbf{q}_j},
\label{eq:Omega5}
\end{equation}
with
\begin{equation}
\begin{aligned}
T_{1}&=\begin{bmatrix} w_{AA} & w_{AB}\\ w_{AB} & w_{AA} \end{bmatrix},\\
T_{2}&=\begin{bmatrix} w_{AA} & w_{AB}\,e^{-2\pi i/3}\\ w_{AB}\,e^{2\pi i/3} & w_{AA} \end{bmatrix},\\
T_{3}&=\begin{bmatrix} w_{AA} & w_{AB}\,e^{2\pi i/3}\\ w_{AB}\,e^{-2\pi i/3} & w_{AA} \end{bmatrix},
\end{aligned}
\label{eq:Tj}
\end{equation}
where $w_{AA},w_{AB}\approx 110$ meV are the physical inter-layer hopping amplitudes, independent of $N_c$ and of the twist angle (but tunable by hydrostatic pressure~\cite{Yankowitz2019}). The off-diagonal phases $e^{\pm 2\pi i/3}$ originate from the AA, AB, and BA stacking points located at the cells $s=1$, $s=n+1$, and $s=2n+1$ along the long diagonal (the Supplemental Material, Sec.~S4). Assembling the diagonal Dirac blocks $h_{\alpha}=-v_F\,\delta\mathbf{k}_{\alpha}\cdot\bm{\sigma}$ and $h_{\beta}=-v_F\,\delta\mathbf{k}_{\beta}\cdot\bm{\sigma}$, Eq.~(\ref{eq:Omega5}) together with Eq.~(\ref{eq:hd}) reproduces the Bistritzer--MacDonald model for two layers AA-stacked before twisting~\cite{Bistritzer2011}.

\subsection{Degeneracy points and energy windows}
\label{sec:degeneracy}

\begin{figure*}[t]
\centering
\includegraphics[width=\textwidth]{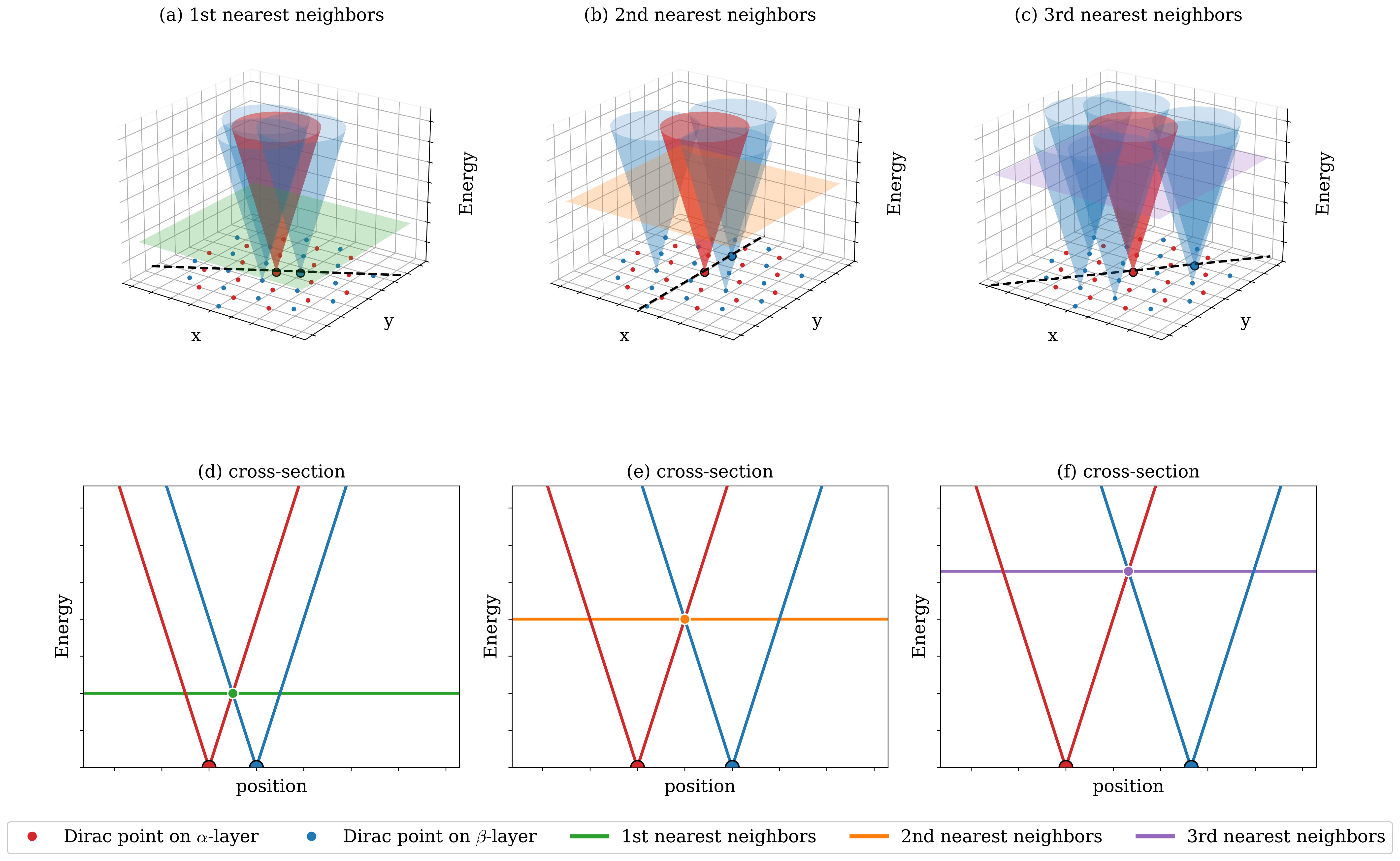}
\caption{Effect of the chosen neighbor order on the degeneracy points, and hence on the energy region described by the resulting effective model.}
\label{fig:003}

\end{figure*}

A defining feature of the present construction is that the degeneracy point at which the perturbative reduction is performed is \emph{not} restricted to the low-energy Dirac cones. Any (quasi-)degeneracy between the two layers---at \emph{any} energy---can be selected, its bands isolated through the block diagonalization, and the inter-layer coupling folded onto that subspace. Which degeneracies actually participate is controlled by the \emph{neighbor order}: retaining only the first shell of momentum transfers couples the nearest Dirac images, keeping higher shells brings in further degeneracies. Each selected degeneracy then produces an effective model valid in the corresponding \emph{energy window} of the spectrum: performing the reduction at the Dirac point yields the low-energy flat bands of the BM model, while performing it at a higher-energy degeneracy yields an effective model for the remote bands, and at a van Hove or $\Gamma$-point degeneracy yet another. In this way a single, uniform procedure generates a hierarchy of effective models, one for each energy region, with the energy of the selected degeneracy as the organizing parameter. This is illustrated schematically in Fig.~\ref{fig:003}, where the neighbor order selects the degeneracy points and thereby the energy region described by the model. 
Expansions performed around the second-nearest coupling or third nearest coupling induced energy degeneracy points can also open band gaps. By appropriately tuning the twist angle, new flat bands can be realized in an analogous manner. In this scenario, the original low-energy degeneracies persist and may couple with these newly emergent flat bands. Our method therefore provides a viable pathway for systematically discovering and engineering new flat-band states in moiré lattice systems.

\subsection{Second-neighbor inter-layer coupling}
\label{sec:secondneighbor}
\begin{figure}[ht]
\centering
\includegraphics[width=\linewidth]{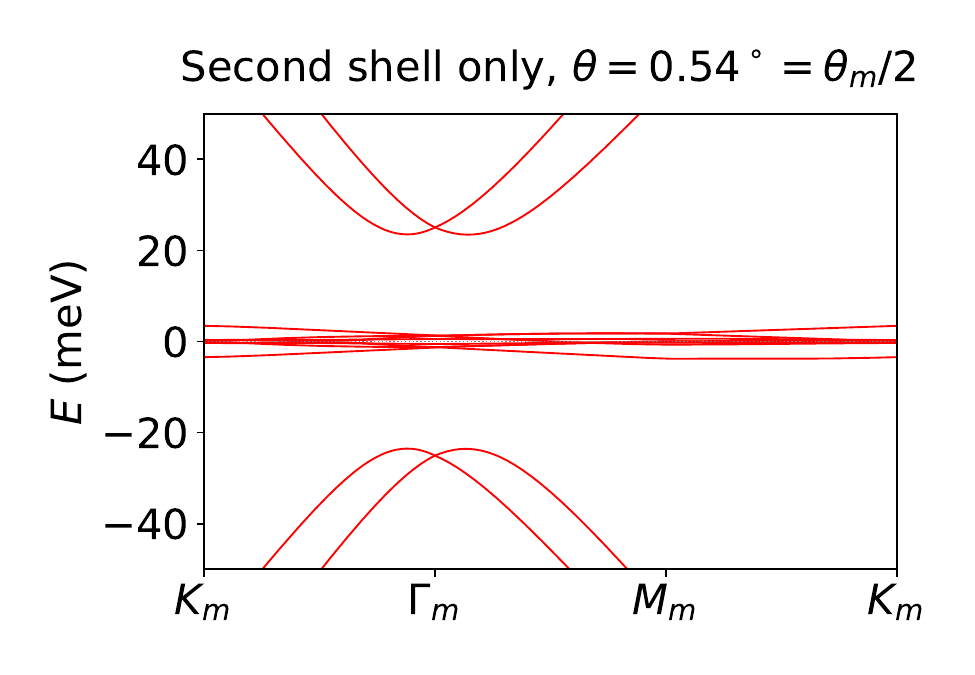}
\caption{Scale self-similarity. Band structure of the second shell alone at
$\theta=0.54^\circ\simeq\theta_m/2$ along $K_m$--$\Gamma_m$--$M_m$--$K_m$: the
two flat bands reappear, confirming that the second shell is the first shell at
the rescaled momentum $q\to2q$. The second shell alone is kept here only to
exhibit this self-similarity; it is not the physical remote band, which requires
both shells and is studied in Sec.~\ref{sec:remote}.}
\label{fig:secondshell}
\end{figure}

The derivation above retained only the nearest-neighbor inter-layer hopping, i.e.\ in the momentum-conservation condition only the three supercell reciprocal vectors $\mathbf{G}$ of minimal length (the first shell). The same procedure extends verbatim to the second-neighbor coupling: in real space this corresponds to inter-layer hopping between atoms displaced by one additional primitive vector, and in reciprocal space to keeping the next three smallest values of $|\mathbf{K}_{\beta}-\mathbf{K}_{\alpha}+\mathbf{G}|$, which are the \emph{doubled} transfers
\begin{equation}
2\mathbf{q}_1,\quad 2\mathbf{q}_2,\quad 2\mathbf{q}_3,
\qquad |2\mathbf{q}_j|=2k_\theta.
\end{equation}
These are \emph{inter-layer} transfers (connecting the $\alpha$-layer Dirac point to the next-nearest $\beta$-layer Dirac images), unlike intra-layer reciprocal vectors such as $\mathbf{q}_i-\mathbf{q}_j$. They yield three additional independent channels,
\begin{equation}
\Omega_{ll'}=\sum_{j=1}^{3} T_j\,\delta_{\delta\mathbf{k}_{\alpha}-\delta\mathbf{k}_{\beta},\;\mathbf{q}_j}
\;+\;\sum_{m=1}^{3} T_m^{(2)}\,\delta_{\delta\mathbf{k}_{\alpha}-\delta\mathbf{k}_{\beta},\;2\mathbf{q}_m},
\end{equation}
where the second-shell matrices are (the Supplemental Material, Sec.~S5)
\begin{equation}
\begin{aligned}
T_1^{(2)}&=\begin{bmatrix} w_{AA}^{(2)} & w_{AB}^{(2)}\\ w_{AB}^{(2)} & w_{AA}^{(2)} \end{bmatrix},\\
T_2^{(2)}&=\begin{bmatrix} w_{AA}^{(2)} & w_{AB}^{(2)}e^{2\pi i/3}\\ w_{AB}^{(2)}e^{-2\pi i/3} & w_{AA}^{(2)} \end{bmatrix},\\
T_3^{(2)}&=\begin{bmatrix} w_{AA}^{(2)} & w_{AB}^{(2)}e^{-2\pi i/3}\\ w_{AB}^{(2)}e^{2\pi i/3} & w_{AA}^{(2)} \end{bmatrix}.
\end{aligned}
\label{eq:Tsecond}
\end{equation}
Since the second-shell transfer $2\mathbf{q}_m$ is parallel to $\mathbf{q}_m$, its azimuthal direction is the same, but the stacking phase is \emph{squared}, so the phase angles become $2\varphi_m=\tfrac{4\pi(m-1)}{3}$. The second shell therefore reproduces the \emph{same} phase set $\{0,\pm 2\pi i/3\}$ as the first shell (with $T_2^{(2)}$ and $T_3^{(2)}$ exchanging roles relative to $T_2,T_3$, since $e^{\pm4\pi i/3}=e^{\mp2\pi i/3}$), introducing no new phase structure but only rescaled amplitudes $w_{AA}^{(2)},w_{AB}^{(2)}$, which are fixed by the same microscopic inter-layer hopping (Slater--Koster $p_z$~\cite{SlaterKoster1954} or DFT tight-binding) evaluated at the second-shell momentum. Because the transfer $2\mathbf{q}_m$ is, to leading order in $\theta$, perpendicular to $\mathbf{K}$, the hopping is evaluated at a momentum whose magnitude exceeds $|\mathbf{K}|$ only at order $\theta^2$; with the parametrization $t(q)=t_0\exp[-\alpha(qd)^\gamma]$ the amplitude ratio is
\begin{equation}
\frac{w^{(2)}}{w}=\frac{t(|\mathbf{K}+2\mathbf{q}_m|)}{t(|\mathbf{K}|)}\simeq 1-\alpha\gamma(|\mathbf{K}|d)^\gamma\,\theta^2\simeq 0.9995,
\end{equation}
so the second shell carries \emph{essentially the same} amplitude as the first shell, the reduction being only about $5\times10^{-4}$. This near-degeneracy of the shell amplitudes---rather than any suppression of the second shell---underlies the scale self-similarity shown in Fig.~\ref{fig:secondshell}: the second shell alone at half the magic angle reproduces the flat bands that the first shell produces at the magic angle, since it is the first shell at the rescaled momentum $q\to2q$.

\subsection{Remote bands and testable predictions}
\label{sec:remote}

\begin{figure*}[t]
\centering
\includegraphics[width=\textwidth]{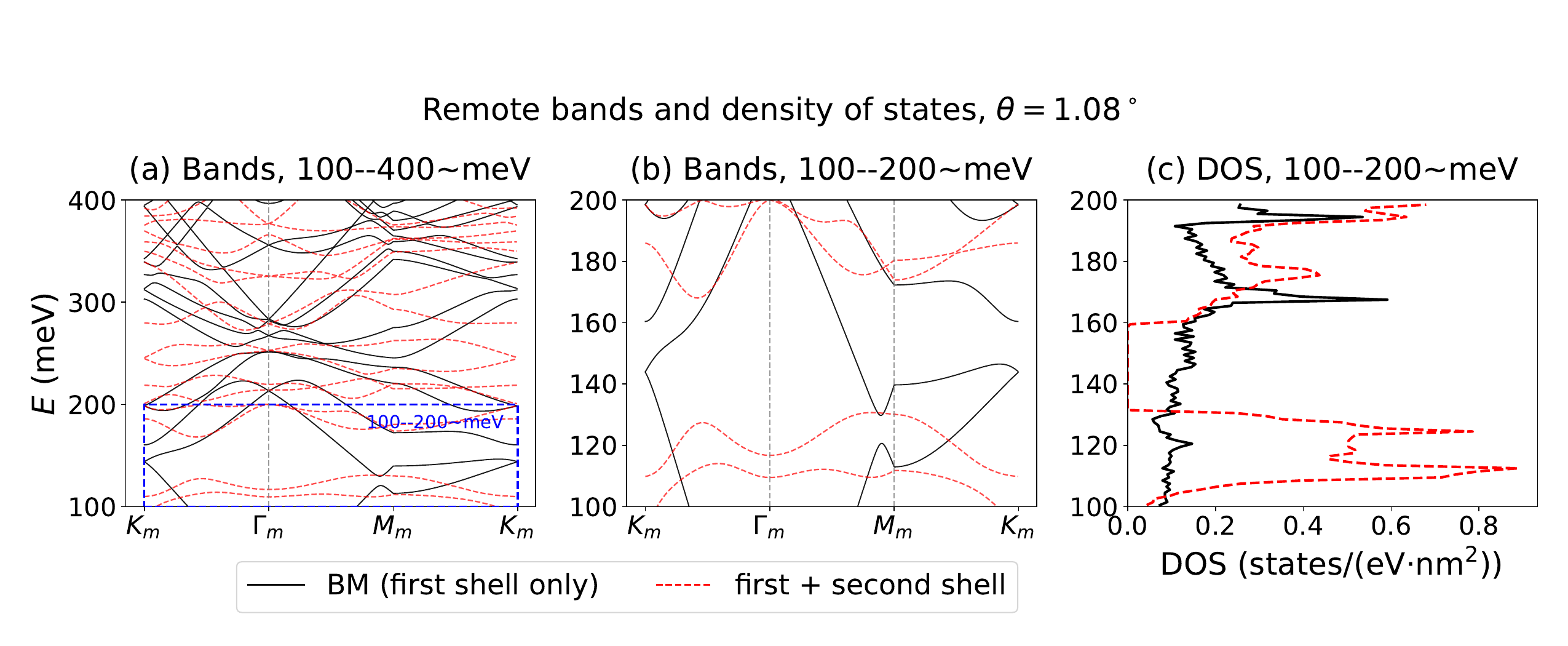}
\caption{Remote bands and density of states at $\theta=1.08^\circ$; black, first
shell only (BM), red dashed, first $+$ second shell. (a) Bands over the
$100$--$400$~meV window; the blue dashed rectangle marks the $100$--$200$~meV
window enlarged in panel (b). (b) Bands over the $100$--$200$~meV window; the
second shell opens the $\Gamma$-point degeneracies retained by BM and rearranges
the remote bands. (c) DOS in the $100$--$200$~meV window; the second shell
produces a sharp van Hove peak near $110$--$120$~meV and opens a gap above it
that the BM DOS lacks.}
\label{fig:remotecompare}
\end{figure*}

The second shell has a decisive consequence for the high-energy sector. Around
charge neutrality the degeneracies involve only the nearest-neighbour Dirac
cones, so the model reduces there to the BM first-shell truncation. Above the
moir\'e $\Gamma$ point, however, six Dirac cones of the Brillouin zone become
simultaneously degenerate---both nearest- and next-nearest---so the first and
second shells enter as first-order terms of the same degenerate perturbation
theory and must be retained together. Keeping the second shell opens the
degeneracies that the BM model leaves intact at the $\Gamma$ point and yields a
faithful description of the remote bands (Fig.~\ref{fig:remotecompare}).

The opening of these degeneracies is imprinted on the density of states. In the
$100$--$200$~meV window the second shell flattens the remote bands and produces
a sharp van Hove singularity near $110$--$120$~meV, and---more decisively---it
opens an energy gap just above this peak that is absent in the BM first-shell
DOS (panel (c) of Fig.~\ref{fig:remotecompare}). The gap is a robust prediction
rather than a finite-size artifact (its convergence with the momentum-network
truncation is discussed in the Supplemental Material, Sec.~\ref{app:conv}).
Detecting this gap, for instance by
re-analyzing existing scanning-tunneling $dI/dV$ spectra, is therefore the most
direct experimental test of the second shell.

\begin{figure}[ht]
\centering
\includegraphics[width=\linewidth]{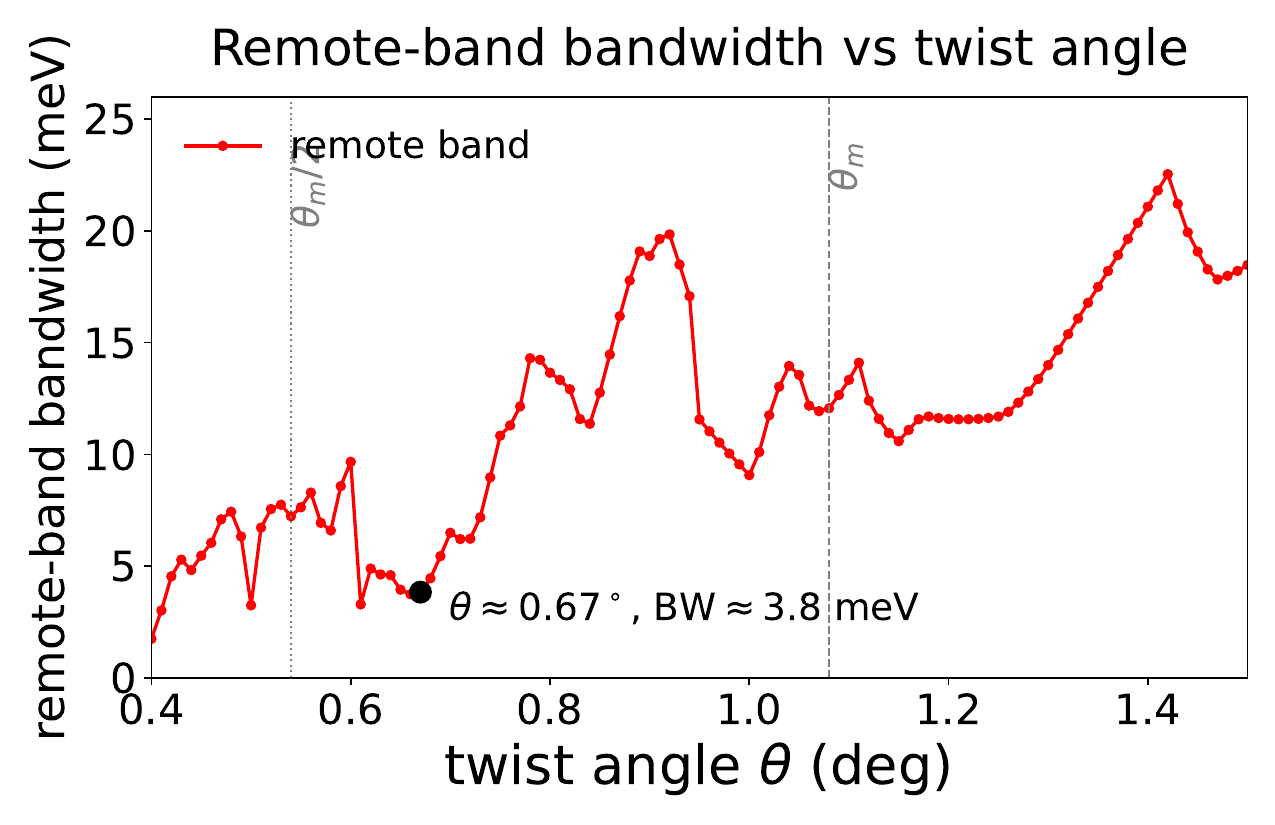}
\caption{Bandwidth of the remote band versus twist angle, tracked at the
second-shell energy window $E\propto\theta$. The bandwidth stays of order
$10$~meV near the magic angle and reaches a minimum of a few meV near
$\theta\simeq0.67^\circ$.}
\label{fig:remotescan}
\end{figure}

The scale self-similarity also governs the twist-angle dependence of the remote
band. Tracking the remote band at its energy window---which scales linearly with
$\theta$ through the degeneracy energy $E_{\rm deg}=v_F|\Delta K|/2$ with
$|\Delta K|=2q\simeq 2k_D\theta$---one finds that its bandwidth stays of order
$10$~meV near the magic angle and reaches a minimum of a few meV near
$\theta\simeq0.67^\circ$ (Fig.~\ref{fig:remotescan}), where the second shell
flattens the remote band into a nearly dispersionless band of width
$\sim4$~meV within the $50$--$100$~meV window; the corresponding band structure
and density of states are shown in the Supplemental Material, Sec.~\ref{app:conv}.

\section{Conclusion}

In summary, this work addresses a long-standing critical issue in moiré physics: conventional theoretical frameworks, exemplified by the classic Bistritzer–MacDonald (BM) continuum model, rely on restrictive low-energy Dirac cone approximations and are only valid for narrow low-energy regimes and small-angle twisted bilayer graphene. These traditional methods fail to provide universal analytical descriptions for strained bilayer lattices, quasiperiodic composite structures, and high-energy band behaviors, leaving a broad class of emergent moiré systems dependent on case-specific numerical simulations and lacking unified physical interpretation. Establishing a general analytical formalism for moiré and quasiperiodic lattices is therefore fundamentally important, as it bridges the gap between structural geometry and full-spectrum electronic properties, enables systematic band engineering, and provides predictive theoretical guidance for designing moiré-based quantum and optoelectronic devices. To resolve these limitations, we develop a universal approximation-free analytical paradigm based on block-circulant diagonalization and quasi-degenerate Löwdin perturbation theory. Different from conventional strategies that start from low-energy Dirac dispersion and expand interlayer couplings around Dirac cone states, our framework adopts an inverse construction scheme: it reduces the full supercell tight-binding Hamiltonian via a 2D-to-1D cyclic supercell mapping and performs perturbative expansion centered on arbitrary energy degeneracy points across the entire energy spectrum, rather than being limited to low-energy Dirac cone couplings. This unique treatment delivers multiple decisive advantages over existing models. First, our method completely abandons conventional low-energy truncation, 
realizing full-energy-spectrum analytical characterization. Second, the degeneracy-centered expansion mechanism breaks the energy and geometric constraints of traditional models, achieving universal applicability for arbitrary commensurate twisted moiré lattices, heterostrained bilayer systems. Third, our framework exhibits excellent self-consistency with classic theories: it rigorously recovers the standard BM model in the low-energy first-shell coupling limit, perfectly reproducing the well-known flat-band physics of magic-angle twisted bilayer graphene. Beyond low-energy regimes, our access to high-energy degeneracy points enables the analytical exploration of previously inaccessible high-energy band features and further predicts higher-order emergent magic angles originating from  high-energy band degeneracies, expanding the family of magic-angle moiré systems beyond the traditional low-energy paradigm. Furthermore, the revealed intrinsic scale self-similarity of moiré lattices provides a unified physical mechanism for understanding hierarchical band regulation in generalized composite lattices. This work eliminates the longstanding lack of universal analytical tools for diverse moiré systems, establishes a general theoretical paradigm for full-energy band analysis, and offers a reliable analytical platform for exploring exotic strongly correlated and topological quantum states in artificial layered lattice materials.

\begin{acknowledgments}

JW was partially supported by the Natural Science Foundation of Guangdong Province (Grant Nos. 2017B030308003, 2019B121203002) and the Science, Technology and Innovation Commission of Shenzhen Municipality (Grant Nos. KYTDPT20181011104202253, JCYJ20170412152620376).
XRW acknowledges the support from  the Guangdong Provincial Quantum Strategy Special Project, the University Development Fund of the Chinese University of Hong Kong, Shenzhen, and 
the National Natural Science Foundation of China (Grants No. 12374122).
\par

The authors acknowledge the use of the AI-based writing tools Doubao and DeepSeek v4 pro for language polishing. The authors take full responsibility for the content of this manuscript.
\par

\emph{Data availability.}$-$The raw data supporting the findings of this manuscript and the custom codes for calculating data are available from the corresponding authors upon reasonable requests. 

\end{acknowledgments}

\section*{Supplemental Material}

\setcounter{section}{0}
\renewcommand{\thesection}{S\arabic{section}}
\renewcommand{\thesubsection}{\thesection.\arabic{subsection}}
\makeatletter
\renewcommand{\p@subsection}{}
\makeatother

\section{Diagonalization of the block-circulant matrix}
\label{app:A}

In this section we derive Eqs.~(\ref{eq:Gamma})--(\ref{eq:omega}) of Section~2 in full detail. Subsection~\ref{app:A1} proves the diagonalization of a circulant matrix by the discrete Fourier transform; Subsection~\ref{app:A2} derives the transformation of the inter-layer coupling block $\omega\to\Omega$ under this diagonalization, which is Eq.~(\ref{eq:omega}).

\subsection{Diagonalization of a circulant matrix}
\label{app:A1}

Consider an $N\times N$ circulant matrix
\begin{equation}
h=\begin{bmatrix}
a_{1} & a_{2} & a_{3} & \cdots & a_{N-1} & a_{N} \\
a_{N} & a_{1} & a_{2} & \cdots & a_{N-2} & a_{N-1} \\
a_{N-1} & a_{N} & a_{1} & \cdots & a_{N-3} & a_{N-2} \\
\vdots & \vdots & \vdots & \ddots & \vdots & \vdots \\
a_{3} & a_{4} & a_{5} & \cdots & a_{1} & a_{2} \\
a_{2} & a_{3} & a_{4} & \cdots & a_{N} & a_{1}
\end{bmatrix},
\label{eq:appcirculant}
\end{equation}
acting on the basis $\langle\Psi|=\langle C_{1}\ C_{2}\ \cdots\ C_{N}|$, so that $H=\langle\Psi|h|\Psi\rangle$. The eigenvalue equation $h|\Psi\rangle=E|\Psi\rangle$ reads, component by component,
\begin{equation}
\begin{aligned}
& a_{1}C_{1}+a_{2}C_{2}+\cdots+a_{N}C_{N}=EC_{1},\\
& a_{N}C_{1}+a_{1}C_{2}+\cdots+a_{N-1}C_{N}=EC_{2},\\
& ...... \\
& a_{2}C_{1}+a_{3}C_{2}+\cdots+a_{1}C_{N}=EC_{N}
\end{aligned}
\end{equation}
In compact form,
\begin{equation}
E\,C_{p}=\sum_{q=1}^{N}a_{q+1-p}\,C_{q},
\label{eq:appeigeneq}
\end{equation}
where the index of $a$ is understood cyclically, $a_{q+1-p}=a_{q+1-p\pm N}$, so that it always lies in $\{1,\dots,N\}$.

We now exploit the translational structure of the circulant matrix by multiplying both sides of Eq.~(\ref{eq:appeigeneq}) by a phase factor $e^{2\pi i(p-1)(l-1)/N}$ with a fixed integer $l\in\{1,\dots,N\}$:
\begin{equation}
e^{\frac{2\pi i(p-1)(l-1)}{N}}\,E\,C_{p}
=e^{\frac{2\pi i(p-1)(l-1)}{N}}\sum_{q=1}^{N}a_{q+1-p}\,C_{q}.
\label{eq:appphase}
\end{equation}
Summing Eq.~(\ref{eq:appphase}) over $p=1,\dots,N$ gives
\begin{equation}
\begin{aligned}
&E\sum_{p=1}^{N}e^{\frac{2\pi i(p-1)(l-1)}{N}}C_{p}\\
&\quad=\sum_{q=1}^{N}C_{q}\Big(\sum_{p=1}^{N}e^{\frac{2\pi i(p-1)(l-1)}{N}}a_{q+1-p}\Big)\\
&\quad=\sum_{q=1}^{N}e^{\frac{2\pi i(q-1)(l-1)}{N}}C_{q}
\Big(\sum_{p=1}^{N}e^{-\frac{2\pi i(q-p)(l-1)}{N}}a_{q+1-p}\Big).
\end{aligned}
\label{eq:appsum}
\end{equation}
The inner sum in the last line is, in fact, independent of $q$. To see this, set $r=q+1-p$, so that $a_{q+1-p}=a_{r}$ and $q-p=r-1$; as $p$ runs over $1,\dots,N$, the cyclic variable $r$ also runs over all of $1,\dots,N$. Hence
\begin{equation}
m_{q}\equiv\sum_{p=1}^{N}e^{-\frac{2\pi i(q-p)(l-1)}{N}}a_{q+1-p}
=\sum_{r=1}^{N}e^{-\frac{2\pi i(r-1)(l-1)}{N}}a_{r},
\label{eq:appmq}
\end{equation}
which is manifestly independent of $q$. We denote this $q$-independent quantity by
\begin{equation}
\Gamma_{l}=\sum_{p=1}^{N}e^{-\frac{2\pi i(p-1)(l-1)}{N}}a_{p},
\label{eq:appGamma}
\end{equation}
where we have renamed the summation index. Substituting Eq.~(\ref{eq:appGamma}) into Eq.~(\ref{eq:appsum}) gives
\begin{equation}
E\sum_{p=1}^{N}e^{\frac{2\pi i(p-1)(l-1)}{N}}C_{p}
=\Gamma_{l}\sum_{q=1}^{N}e^{\frac{2\pi i(q-1)(l-1)}{N}}C_{q}.
\label{eq:appreduced}
\end{equation}
Defining the linear combination
\begin{equation}
x_{l}=\sum_{p=1}^{N}e^{\frac{2\pi i(p-1)(l-1)}{N}}C_{p},
\label{eq:appx}
\end{equation}
Eq.~(\ref{eq:appreduced}) reduces to the scalar eigenvalue equation
\begin{equation}
E\,x_{l}=\Gamma_{l}\,x_{l}.
\end{equation}
Since this holds for every $l=1,\dots,N$, the circulant matrix $h$ is diagonalized into
\begin{equation}
\widetilde{h}=\mathrm{diag}(\Gamma_{1},\Gamma_{2},\dots,\Gamma_{N}),
\end{equation}
with the (normalized) new basis $\langle\Psi_{X}|=\langle X_{1}\ X_{2}\ \cdots\ X_{N}|$ where
\begin{equation}
X_{l}=\frac{1}{\sqrt{N}}x_{l}
=\frac{1}{\sqrt{N}}\sum_{p=1}^{N}e^{\frac{2\pi i(p-1)(l-1)}{N}}C_{p}.
\label{eq:appX}
\end{equation}
This is precisely the discrete Fourier transform: Eq.~(\ref{eq:appGamma}) is Eq.~(\ref{eq:Gamma}) and Eq.~(\ref{eq:appX}) is Eq.~(\ref{eq:cx}) of the main text. The same derivation applies verbatim when each $a_{p}$ is itself a Hermitian matrix (the block-circulant case), in which case $\Gamma_{l}$ is a matrix.

\subsection{Folded inter-layer coupling}
\label{app:A2}

We now derive Eq.~(\ref{eq:omega}), the transformation of the inter-layer block under the above diagonalization. Suppose the two layers contain $M$ and $N$ atoms, respectively, so that the intra-layer Hamiltonian is block-diagonal in the $X$-basis,
\begin{equation}
H_{0}=\begin{bmatrix} H_{M\times M}^{\alpha} & 0 \\ 0 & H_{N\times N}^{\beta} \end{bmatrix},
\end{equation}
The inter-layer coupling in the original cell basis is an $M\times N$ matrix $V_{0}$ with entries
\begin{equation}
[V_{0}]_{pq}=\omega_{pq}=t_{pq}\,e^{-i\mathbf{k}\cdot\Delta_{pq}},
\end{equation}
where $t_{pq}$ is the hopping between cell $p$ of layer $\alpha$ and cell $q$ of layer $\beta$, and $\Delta_{pq}$ their in-plane separation. The two bases are related by
\begin{equation}
\begin{aligned}
  & X_{x}^{\alpha}=\frac{1}{\sqrt{M}}\sum_{p=1}^{M}e^{\frac{2\pi i(p-1)(x-1)}{M}}C_{p}^{\alpha},\\
  & X_{y}^{\beta}=\frac{1}{\sqrt{N}}\sum_{q=1}^{N}e^{\frac{2\pi i(q-1)(y-1)}{N}}C_{q}^{\beta}.
\end{aligned}
\end{equation}
Writing the inter-layer term in both bases,
\begin{equation}
\begin{aligned}
&\frac{1}{\sqrt{MN}}\sum_{p,q}\sum_{x,y}\Omega_{xy}
e^{2\pi i\left(\frac{(p-1)(x-1)}{M}-\frac{(q-1)(y-1)}{N}\right)}C_{p}^{\alpha\,\dagger}C_{q}^{\beta}\\
&\qquad=\sum_{p,q}\omega_{pq}\,C_{p}^{\alpha\,\dagger}C_{q}^{\beta},
\end{aligned}
\label{eq:appinterbase}
\end{equation}
and equating coefficients of $C_{p}^{\alpha\,\dagger}C_{q}^{\beta}$ for each pair $(p,q)$, we obtain
\begin{equation}
\frac{1}{\sqrt{MN}}\sum_{x,y}\Omega_{xy}
e^{2\pi i\left(\frac{(p-1)(x-1)}{M}-\frac{(q-1)(y-1)}{N}\right)}=\omega_{pq}.
\label{eq:appcoeff}
\end{equation}
The solution of Eq.~(\ref{eq:appcoeff}) is the inverse discrete Fourier transform,
\begin{equation}
\Omega_{xy}=\sum_{p,q}\frac{1}{\sqrt{MN}}
e^{-2\pi i\left(\frac{(p-1)(x-1)}{M}-\frac{(q-1)(y-1)}{N}\right)}\omega_{pq},
\label{eq:appOmega}
\end{equation}
which is Eq.~(\ref{eq:omega}). To verify that Eq.~(\ref{eq:appOmega}) indeed solves Eq.~(\ref{eq:appcoeff}), substitute it into the left-hand side of Eq.~(\ref{eq:appcoeff}):
\begin{equation}
\frac{1}{MN}\sum_{x,y}\sum_{s_{\alpha},s_{\beta}}
e^{2\pi i\left(\frac{(x-1)(p-s_{\alpha})}{M}-\frac{(y-1)(q-s_{\beta})}{N}\right)}\omega_{s_{\alpha}s_{\beta}}.
\label{eq:appsubst}
\end{equation}
When $s_{\alpha}=p$ and $s_{\beta}=q$, the exponential is unity and the double sum over $(x,y)$ contributes a factor $MN$, giving the single term $\omega_{pq}$. When $s_{\alpha}\neq p$ or $s_{\beta}\neq q$, the sum vanishes: for instance, if $s_{\beta}\neq q$, then
\begin{equation}
\sum_{y=1}^{N}e^{-2\pi i\frac{(y-1)(q-s_{\beta})}{N}}
=\sum_{j=0}^{N-1}\zeta^{j},
\qquad
\zeta\equiv e^{-2\pi i\frac{q-s_{\beta}}{N}}\neq 1,
\end{equation}
and the geometric series gives
\begin{equation}
\sum_{j=0}^{N-1}\zeta^{j}=\frac{1-\zeta^{N}}{1-\zeta}=0,
\end{equation}
because $\zeta^{N}=e^{-2\pi i(q-s_{\beta})}=1$. The same argument applies to the $x$-sum when $s_{\alpha}\neq p$. Hence Eq.~(\ref{eq:appsubst}) equals $\omega_{pq}$, proving that Eq.~(\ref{eq:appOmega}) is the correct folded inter-layer block.

\section{Derivation of the 1D-chain effective Hamiltonian}
\label{app:B}
We detail the derivation of Eq.~(\ref{eq:1dchain}). Consider two chains $\alpha$ and $\beta$ with periods $r_{\alpha}=1$ and $r_{\beta}=4/3$, so that the supercell period is $R=4$, containing four $\alpha$-atoms and three $\beta$-atoms. The single-chain hopping $t_{\alpha}$ ($t_{\beta}$) produces the intra-chain blocks; their circulant eigenvalues follow from Sec.~S1 as
\begin{equation}
\begin{aligned}
&\Gamma_{\alpha,i}=2t_{\alpha}\cos\Big(\mathbf{k}\mathbf{r}_{\alpha}+\frac{2(i-1)\pi}{4}\Big),\\
&\Gamma_{\beta,i}=2t_{\beta}\cos\Big(\mathbf{k}\mathbf{r}_{\beta}+\frac{2(i-1)\pi}{3}\Big).
\end{aligned}
\end{equation}
For the inter-chain part we retain only the hopping between atoms at coincident in-plane positions, $t_{\alpha,\beta}$; the folded matrix element follows from Eq.~(\ref{eq:omega}) with $N_{1}=4$, $N_{2}=3$ and $\omega_{s_{\alpha}s_{\beta}}=t_{\alpha,\beta}\delta_{s_{\alpha},s_{\beta}}$, giving the constant $\Omega=\tfrac{1}{\sqrt{12}}t_{\alpha,\beta}$ that couples every coincident pair. Since the inter-chain coupling is weak, $t_{\alpha,\beta}\ll t_{\alpha},t_{\beta}$, first-order degenerate perturbation theory (Section~2) confines the effect to the band degeneracies; selecting the degeneracies inside the dashed window of Fig.~\ref{fig:1dchain} and reading off which chain bands participate in each, one obtains the $7\times7$ Hamiltonian of Eq.~(\ref{eq:1dchain}) in the ordered basis $[X^{\alpha}_{1},X^{\beta}_{2},X^{\beta}_{3},X^{\alpha}_{3},X^{\alpha}_{4},X^{\alpha}_{2},X^{\beta}_{1}]$.

\section{Derivation of continuum interlayer coupling for 1D bilayer moir\'e chain}
\label{app:C}

We consider a one-dimensional bilayer lattice. The upper $\alpha$-layer has lattice constant $a_1=1$ with sites $x_n=n$ and nearest-neighbor hopping $t_\alpha$. The lower $\beta$-layer has lattice constant $a_2=\alpha_0$ with sites $y_m=m\alpha_0$ and hopping $t_\beta$. The parameter $\alpha_0$ is irrational, and we adopt the $k$-th continued-fraction rational approximation $p_k/q_k \to \alpha_0$ with system sizes $N_1=p_k$, $N_2=q_k$.

The interlayer hopping is defined with spatial cutoff $\delta$:
\begin{equation}
\omega_{nm}=
\begin{cases}
t_\perp, & |n-m\alpha_0|<\delta\\
0, & \text{otherwise}
\end{cases}.
\label{eq:match-condition}
\end{equation}
Define matching set $\mathcal M(n)=\{m\mid |n-m\alpha_0|<\delta\}$, restricted to at most two matched sites $m_0(n)$ (primary) and $m_0(n)+1$ (secondary), with indicator functions $\eta_0(n),\eta_1(n)$. The discrete folded interlayer coupling reads
\[
\Omega_{xy}^{(k)}
=\frac{t_\perp}{\sqrt{N_1 N_2}}
\sum_{n=1}^{N_1}\sum_{m\in \mathcal M(n)}
e^{\left\{
-i\left[
k_x^{(k)}(n-1)
- k_y^{(k)}(m-1)
\right]
\right\}},
\]
where $k_x^{(k)}=2\pi(x-1)/N_1$, $k_y^{(k)}=2\pi(y-1)/N_2$.
Introduce phase kernel
\[
U_{k,n}(y)
=\eta_0(n)e^{+i k_y^{(k)}(m_0(n)-1)}
+\eta_1(n)e^{+i k_y^{(k)} m_0(n)},
\]
then
\[
\Omega_{xy}^{(k)}
=\frac{t_\perp \,e^{ik_x^{(k)}}}{\sqrt{N_1 N_2}}
\sum_{n=1}^{N_1}
e^{-i k_x^{(k)} n}\; U_{k,n}(y).
\]

In the high-order limit $k\to\infty$, $m_0(n)\approx n/\alpha_0$. Define effective weight
\[
\widetilde W_k(n)=\eta_0(n)+\eta_1(n)e^{ik_y^{(k)}}.
\]
The asymptotic form becomes
\begin{multline}
\Omega_{xy}^{(k)}
\approx
\frac{t_\perp e^{i\left(k_x^{(k)}-k_y^{(k)}\right)}}{\sqrt{N_1N_2}}
\sum_{n=1}^{N_1}\widetilde W_k(n)\\
\times \exp\left[
-i\left(k_x^{(k)}-\frac{k_y^{(k)}}{\alpha_0}\right)n
\right].
\end{multline}

Take continuum limit $k\to\infty$ and  map the discrete lattice index $n$ to a continuous coordinate $z\equiv n$: $k_x^{(k)}\to k_x$, $k_y^{(k)}\to k_y$, $\sum_n \to \int dz$,  $\eta_0(n) \to \eta_0(z)$,  $\eta_1(n) \to \eta_1(z)$,and $\widetilde W_k(n)\to \widetilde W(z)$. The dominant Fourier components of $\widetilde W(z)$ are $\pm q$ with moir\'e wavevector
\begin{equation}
q=2\pi\left(1-\frac{1}{\alpha_0}\right),\qquad
\widetilde W(z)=C_+ e^{iqz}+C_- e^{-iqz}.
\label{eq:W}
\end{equation}

We prove the above equation in the followings. To analyze the periodic structure of these matching functions, we define the lattice misalignment phase
\begin{equation}\label{eq:phase-def}
\phi(z) = q z=2\pi z\left(1-\frac{1}{\alpha_0}\right),
\end{equation}
which quantifies the relative spatial slip between the $\alpha$-layer and $\beta$-layer as a function of the continuous coordinate $z$. The interlayer distance can be rewritten modulo lattice translation:
\[
n-m\alpha_0 =n\left(1-\frac{m}{n}\right)\approx  n\left(1-\frac{1}{\alpha_0}\right)\equiv \frac{\phi(z)}{2\pi} \pmod{1}.
\]
The above approximation is token in the continuum limit $k\to \infty$. This identity demonstrates that the matching condition \eqref{eq:match-condition} depends exclusively on the phase $\phi(z)$ modulo $2\pi$, rather than on the absolute coordinate $z$.
As a consequence, the indicator functions can be expressed as pure functions of the misalignment phase,
\begin{equation}\label{eq:eta-phase-function}
\eta_{0,1}(z) = F_{0,1}(\phi(z)),
\end{equation}
where $F_0(\phi)$ and $F_1(\phi)$ are inherently $2\pi$-periodic functions satisfying
\[
F_{0,1}(\phi+2\pi) = F_{0,1}(\phi).
\]

We now introduce the fundamental moir\'e wavevector as above and the corresponding moir\'e wavelength
\begin{equation}\label{eq:moire-param}
\qquad L_M = \frac{1}{1-1/\alpha_0}.
\end{equation}
Performing a spatial translation by one full moir\'e period $z\to z+L_M$, the misalignment phase transforms as
\[
\phi(z+L_M) = q (z+L_M) = \phi(z) + 2\pi.
\]
The term $-2\pi$ is a trivial integer multiple of the fundamental phase period, which vanishes under modulo-$2\pi$ equivalence.
We thus obtain the key relation
\begin{equation}\label{eq:phase-quasiperiod}
\phi(z+L_M) \equiv \phi(z) \pmod{2\pi}.
\end{equation}

Substituting into the phase representation of the indicator functions, we arrive the following,
\[
\eta_i(z+L_M) = F_i(\phi(z+L_M)) = F_i(\phi(z)) = \eta_i(z)\ (i=0,1).
\]
Although the irrational ratio $\alpha_0$ forbids strict translational periodicity on the original lattice scale, the matching functions exactly repeat themselves upon translation by the moir\'e wavelength $L_M$ in the phase-modulo sense.
Therefore, both $\eta_0(z)$ and $\eta_1(z)$ are rigorously quasiperiodic functions modulated by the fundamental moir\'e wavevector $q$. So Eq.(\ref{eq:W}) is proved.

Substitution of $\tilde{W}$ into $\Omega_{xy}^{k}$ yields
\begin{multline}
\Omega(k_x,k_y) \propto t_\perp C_+ \delta\left(k_x-\frac{k_y}{\alpha_0}-q\right)\\
+ t_\perp C_- \delta\left(k_x-\frac{k_y}{\alpha_0}+q\right).
\end{multline}
The \textbf{resonance condition} for nonvanishing interlayer hopping is
\[
k_x-\frac{k_y}{\alpha_0}=\pm q.
\]

Combining constants into a complex renormalized amplitude $A$, the real-space interlayer multiplication operator reads
\[
V(x)=A\,t_\perp \cos(qx).
\]
The full continuum two-band Hamiltonian is
\[
H_{\rm MAM}=
\begin{pmatrix}
-t_\alpha\partial_x^2 & A t_\perp \cos(qx)\\
A^* t_\perp \cos(qx) & -t_\beta\partial_x^2
\end{pmatrix}.
\]
Retaining only primary and secondary matched sites renormalizes $A$ but does not generate higher harmonics such as $\cos(2qx)$.


\section{Two‑Chain Quasiperiodic Model: Momentum‑Space Two‑Component Recursion Relations and the Infinite‑Dimensional Linear System}

We perform the standard Bloch transformation
\begin{equation}
c_{\alpha,n} = \frac{1}{\sqrt{N}}\sum_{k_1} e^{ik_1 n}\psi_{\alpha,k_1},\quad
c_{\beta,n} = \frac{1}{\sqrt{N}}\sum_{k_2} e^{ik_2 n}\psi_{\beta,k_2}.
\end{equation}
and obtain the Hamiltonian of the above system in momentum space,
\begin{widetext}
\begin{equation}
 H_{\rm MAM}=
\begin{pmatrix}
-2t_\alpha\cos k_1  & \frac{A t_\perp}{2}\left[\delta_{k_1-k_2,+q}+ \delta_{k_1-k_2,-q}\right]\\
\frac{A t_\perp}{2}\left[\delta_{k_1-k_2,+q}+ \delta_{k_1-k_2,-q}\right] & -2t_\beta\cos k_2
\end{pmatrix}. 
\end{equation}
\end{widetext}

Since $q$ is irrational, the set of wave‑vectors $\{k+mq\mid m\in\mathbb{Z}\}$ densely fills the Brillouin zone $[0,2\pi)$, producing an infinite tower of coupled momentum states.
Define momentum‑space wave‑function components indexed by integer $m$:
\begin{equation}
\phi_m = \psi_\alpha(k+mq),\quad \chi_m = \psi_\beta(k+mq),\quad m\in\mathbb{Z}.
\end{equation}
and denote $\varepsilon_{\alpha\beta}(k)=-2t_{\alpha,\beta}\cos(k)$.

The factors $e^{\pm iqn}$ appearing in the inter‑layer coupling act as momentum shift operators $k\to k\pm q$. Substituting the Bloch expansion yields a system of coupled second‑order linear difference equations for the momentum‑space amplitudes:
\begin{equation}
\begin{cases}
\big(\varepsilon_\alpha(k+mq)-E\big)\phi_m + \dfrac{A t_\perp}{2}\big(\chi_{m-1}+\chi_{m+1}\big) = 0,\\[4pt]
\big(\varepsilon_\beta(k+mq)-E\big)\chi_m + \dfrac{A t_\perp}{2}\big(\phi_{m-1}+\phi_{m+1}\big) = 0.
\end{cases}
\label{eq:recursion_2comp}
\end{equation}

The second‑order recursion can be converted into a first‑order iterative map by introducing the four‑dimensional state vector
\begin{equation}
\Psi_m =
\begin{pmatrix}
\phi_m\\
\chi_m\\
\phi_{m+1}\\
\chi_{m+1}
\end{pmatrix}.
\end{equation}
There exists a $4\times4$ transfer matrix $M_m$ such that
\begin{equation}
\Psi_{m+1} = M_m \Psi_m,\quad m\in\mathbb{Z},
\end{equation}
with explicit matrix form
\begin{widetext}
\begin{equation}
M_m =
\begin{pmatrix}
0 & 0 & 1 & 0\\
0 & 0 & 0 & 1\\[4pt]
-1 & 0 & 0 & -\dfrac{2(\varepsilon_\beta(k+mq)-E)}{At_\perp}\\[4pt]
0 & -1  & -\dfrac{2(\varepsilon_\alpha(k+mq)-E)}{A t_\perp} & 0
\end{pmatrix}.
\label{eq:transfer_matrix}
\end{equation}
\end{widetext}

The index $m$ runs over all integers, so $\{\phi_m,\chi_m\}$ constitutes an infinite set of unknown amplitudes. The recursion~\eqref{eq:recursion_2comp} is therefore equivalent to an infinite‑dimensional homogeneous eigenvalue problem
\begin{equation}
\mathbf{M}\cdot\boldsymbol{\Psi}=E\boldsymbol{\Psi},
\end{equation}
where $\mathbf{M}$ is an infinite, tridiagonal quasiperiodic matrix with non‑vanishing matrix elements only between sites $m$ and $m\pm1$.

Because $q$ is irrational, the shifted dispersion functions $\varepsilon_\alpha(k+mq)$ and $\varepsilon_\beta(k+mq)$ never repeat periodically as $m$ increases. Consequently the operator cannot be reduced to a finite‑size matrix, and no elementary closed‑form dispersion relation exists.

The above infinite momentum‑space recursion is strictly dual to the original real‑space two‑chain quasiperiodic problem:
\begin{itemize}
\item real‑space lattice index $n\;\longleftrightarrow\;$ momentum‑space index $m$,
\item real‑space quasiperiodic potential $\cos(qn)\;\longleftrightarrow\;$ momentum‑space quasiperiodic dispersion $\varepsilon_{\alpha/\beta}(k+mq)$,
\item real‑space localized eigenstate $\;\longleftrightarrow\;$ extended momentum‑space wave‑function,
\item real‑space extended eigenstate $\;\longleftrightarrow\;$ localized momentum‑space wave‑function.
\end{itemize}

For irrational modulation $q$, the spectrum of this infinite‑dimensional quasiperiodic operator forms a Cantor set. The practical numerical diagonalization strategy proceeds as follows:
\begin{enumerate}
\item Take Pell rational approximants $\alpha_k=p_k/q_k$ to $\sqrt{2}$ and select a finite supercell of size $N=q_k$, truncating the infinite system into a $2N\times 2N$ finite‑dimensional matrix.
\item Increase the index $k$ (and hence the supercell size) sequentially and diagonalize the truncated matrix to obtain energy eigenvalues.
\item As the supercell grows, the original bands undergo repeated fragmentation with infinitely many gaps opening, converging toward a Cantor‑set spectrum.
\item Compute the box‑counting fractal dimension to verify the fractal nature of the spectrum.
\end{enumerate}

The asymptotic Lyapunov exponent $\gamma$ quantifies exponential growth or decay of the momentum‑space wave‑function:
\begin{equation}
\gamma = \lim_{L\to\infty}\frac{1}{L}\ln\left\| \prod_{m=1}^L M_m(E) \right\|.
\label{eq:lyapunov_def}
\end{equation}
The physical interpretation, dictated by real‑momentum duality, is:
\begin{itemize}
\item $\gamma>0$: momentum‑space wave‑function exponentially localized $\Longleftrightarrow$ real‑space extended itinerant state,
\item $\gamma=0$: critical multifractal state,
\item $\gamma<0$: momentum‑space wave‑function extended $\Longleftrightarrow$ real‑space exponentially localized state.
\end{itemize}
Numerically, one multiplies successive transfer matrices with periodic Gram--Schmidt re‑orthogonalization to suppress numerical overflow. Scanning over energy $E$ yields the function $\gamma(E)$; energies satisfying $\gamma(E_c)=0$ define mobility‑edge locations.

The with varying coupling $t_\perp$, the system show different phases such as locilized phase and extended phase. Critical‑coupling is summarized as followings, 
\begin{enumerate}
\item Symmetric case $t_\alpha=t_\beta=t$: the exact critical inter‑chain coupling is $\boldsymbol{t_{\perp c}=2t}$.
\item Asymmetric case $t_\alpha\ne t_\beta$: no universal exact critical point exists, mobility edges emerge inside the spectrum, and the characteristic approximate threshold obeys $\boldsymbol{t_{\perp c}\approx 2\sqrt{t_\alpha t_\beta}}$.
\item The whole problem belongs to the spectral theory of infinite‑dimensional quasiperiodic linear operators. Reliable solutions require rational‑approximant numerics, the spectrum is Cantor‑like, and the localization‑delocalization transition is determined by the sign of the Lyapunov exponent.

\end{enumerate}

\section{Commensurate geometry of twisted bilayer graphene}
\label{app:D}
Here we assemble the geometric ingredients of Section~4.2. With $\mathbf{a}_{1}=a(1,0)$ and $\mathbf{a}_{2}=a(\tfrac12,\tfrac{\sqrt3}{2})$, and rotation $R_{\theta}\approx I+\theta R_{\pi/2}$, the commensurate supercell satisfies
\begin{equation}
\mathbf{A}_{1}=m\mathbf{a}_{1}+(n+1)\mathbf{a}_{2}=(I+\theta R_{\pi/2})(m\mathbf{a}_{1}+n\mathbf{a}_{2}),
\end{equation}
Here, without loss of generality, we assume by rotation a small angle, a site $(m,n)$ in the second layer is rotated to the site $(m,n+1)$ on the first layer.  This together with $\mathbf{a}_{2}=R_{\pi/3}\mathbf{a}_{1}$, gives $m=-2n$ and $\theta=\sqrt{3}/(3n)$. Thus $\mathbf{A}_{1}=-2n\mathbf{a}_{1}+(n+1)\mathbf{a}_{2}$ and $\mathbf{A}_{2}=R_{\pi/3}\mathbf{A}_{1}=-(n+1)\mathbf{a}_{1}+(-n+1)\mathbf{a}_{2}$, with $N_{c}=3n^{2}+1$. The chain factor $f$ is obtained from the modular condition $[1+\alpha(n+1)+\beta(-n+1)]\bmod N_{c}=0$; the solution $\alpha=\beta=\tfrac{3n^{2}}{2}$ (valid for even $n$, which also ensures $\gcd(n+1,-n+1)=1$) yields $f=\tfrac32n(n+1)+1$. The supercell reciprocal vectors, dual to $\mathbf{A}_{1},\mathbf{A}_{2}$, are
\begin{equation}
\begin{aligned}
& \mathbf{B}_{1}=\frac{2\pi}{aN_{c}}\Big(1-n,\frac{3n+1}{\sqrt3}\Big),\\
& \mathbf{B}_{2}=\frac{2\pi}{aN_{c}}\Big(-(1+n),\frac{1-3n}{\sqrt3}\Big).
\end{aligned}
\end{equation}
Since $\mathbf{B}_{1}$ generates the cyclic quotient group, $\mathbf{B}_{2}\equiv c\,\mathbf{B}_{1}\pmod{\mathbf{b}_{1},\mathbf{b}_{2}}$ with $c\equiv-(n+1)(-n+1)^{-1}\pmod{N_{c}}$.

\emph{Remark on the modular inverse.} The symbol $(-n+1)^{-1}$ denotes the multiplicative inverse of $-n+1$ modulo $N_{c}$, an integer of order $N_{c}$; it must not be approximated by $-1/n$. For instance, at $n=2$ ($N_{c}=13$), $(-n+1)^{-1}=(-1)^{-1}\equiv 12\pmod{13}$; at $n=4$ ($N_{c}=49$), $(-n+1)^{-1}=(-3)^{-1}\equiv 16\pmod{49}$. These integers are unrelated to the real number $1/n$.

\section{Derivation of the inter-layer block for TBG}
\label{app:E}
We derive Eq.~(\ref{eq:Omega5}) without presupposing momentum conservation. The
folded interlayer block $\Omega^{ab}_{ll'}(\mathbf{k}_{\alpha},\mathbf{k}_{\beta})$ is
evaluated with independent layer momenta $\mathbf{k}_{\alpha}$, $\mathbf{k}_{\beta}$
and folding vectors $\mathbf{G}_{l}=(l-1)\mathbf{B}_{1}$,
$\mathbf{G}'_{l'}=(l'-1)\mathbf{B}'_{1}$. It carries the Bloch phase of the two
layers together with the folding phase,
\begin{multline}
  \Omega^{ab}_{ll'}(\mathbf{k}_{\alpha},\mathbf{k}_{\beta})
  =\sum_{s_{\alpha},s_{\beta}}\frac{1}{\sqrt{N_{\alpha}N_{\beta}}}\,
  t(\mathbf{r}_{s_{\alpha}}+\bm{\tau}_{a}-\mathbf{r}_{s_{\beta}}-\bm{\tau}_{b})\\
  \times e^{-i[\mathbf{k}_{\alpha}\cdot(\mathbf{r}_{s_{\alpha}}+\bm{\tau}_{a})
  -\mathbf{k}_{\beta}\cdot(\mathbf{r}_{s_{\beta}}+\bm{\tau}_{b})]}\\
  \times e^{-i[\mathbf{G}_{l}\cdot\mathbf{r}_{s_{\alpha}}
  -\mathbf{G}'_{l'}\cdot\mathbf{r}_{s_{\beta}}]},
\end{multline}
with no relation between $\mathbf{k}_{\alpha}$ and $\mathbf{k}_{\beta}$ assumed.
Combining the two phases into a single plane-wave phase at the unfolded momentum,
\begin{multline}
  e^{-i[\mathbf{k}_{\alpha}\cdot(\mathbf{r}_{s_{\alpha}}+\bm{\tau}_{a})
  -\mathbf{k}_{\beta}\cdot(\mathbf{r}_{s_{\beta}}+\bm{\tau}_{b})]}\,
  e^{-i[\mathbf{G}_{l}\cdot\mathbf{r}_{s_{\alpha}}
  -\mathbf{G}'_{l'}\cdot\mathbf{r}_{s_{\beta}}]}\\
  =e^{-i[(\mathbf{k}_{\alpha}+\mathbf{G}_{l})\cdot(\mathbf{r}_{s_{\alpha}}+\bm{\tau}_{a})
  -(\mathbf{k}_{\beta}+\mathbf{G}'_{l'})\cdot(\mathbf{r}_{s_{\beta}}+\bm{\tau}_{b})]}\\
  \times e^{i[\mathbf{G}_{l}\cdot\bm{\tau}_{a}
  -\mathbf{G}'_{l'}\cdot\bm{\tau}_{b}]},
\end{multline}
and substituting the Fourier representation
$t(\mathbf{r})=\int\frac{d^{2}p}{(2\pi)^{2}}\,\tilde{t}(\mathbf{p})\,e^{i\mathbf{p}\cdot\mathbf{r}}$,
the two cell sums are carried out with the Poisson identity
\begin{equation}
  \sum_{s=1}^{N}e^{i\mathbf{x}\cdot(s-1)\mathbf{a}_{1}}
  =N\sum_{\mathbf{G}}\delta_{\mathbf{x},\mathbf{G}},
\end{equation}
giving
\begin{multline}
  \Omega^{ab}_{ll'}(\mathbf{k}_{\alpha},\mathbf{k}_{\beta})
  =\frac{1}{\sqrt{A_{\alpha}A_{\beta}}}\sum_{\mathbf{G}_{\alpha},\mathbf{G}_{\beta}}\\
  \times e^{i[(\mathbf{G}_{l}+\mathbf{G}_{\alpha})\cdot\bm{\tau}_{a}
  -(\mathbf{G}'_{l'}+\mathbf{G}_{\beta})\cdot\bm{\tau}_{b}]}\\
  \times \tilde{t}(\mathbf{k}_{\alpha}+\mathbf{G}_{l}+\mathbf{G}_{\alpha})\,
  \delta_{\mathbf{k}_{\alpha}+\mathbf{G}_{l}+\mathbf{G}_{\alpha},\mathbf{k}_{\beta}+\mathbf{G}'_{l'}+\mathbf{G}_{\beta}},
\end{multline}
where $A_{\alpha},A_{\beta}$ are the primitive-cell areas. The Kronecker delta,
which arises \emph{from} the cell sums, is the momentum conservation---it is
obtained, not assumed. It is the generalized umklapp condition
$\mathbf{k}_{\alpha}+\mathbf{G}_{l}+\mathbf{G}_{\alpha}
=\mathbf{k}_{\beta}+\mathbf{G}'_{l'}+\mathbf{G}_{\beta}$.

We now specialize to the low-energy Dirac points. Each folded band is expanded
around an unfolded Dirac cone $\mathbf{K}_{\lambda}^{0}$; the primitive momentum is
$\mathbf{k}_{\lambda}=\mathbf{K}_{\lambda}^{0}+\delta\mathbf{k}_{\lambda}$. Inserting
this into the delta function gives the momentum-conservation condition
\begin{equation}
  \delta\mathbf{k}_{\alpha}-\delta\mathbf{k}_{\beta}
  =\mathbf{K}_{\beta}^{0}+\mathbf{G}'_{l'}+\mathbf{G}_{\beta}
  -\mathbf{K}_{\alpha}^{0}-\mathbf{G}_{l}-\mathbf{G}_{\alpha}.
\end{equation}
The lowest-energy window is selected by requiring the two Dirac cones to be
separated by one of the three moir\'e wave vectors,
\begin{equation}
  (\mathbf{K}_{\alpha}^{0}+\mathbf{G}_{l})-(\mathbf{K}_{\beta}^{0}+\mathbf{G}'_{l'})
  =\mathbf{q}_{i},\qquad i=1,2,3,
\end{equation}
so that the $\alpha$-layer Dirac cone couples to the three $\beta$-layer Dirac
cones, $\delta\mathbf{k}_{\beta}=\delta\mathbf{k}_{\alpha}+\mathbf{q}_{j}$
($j=1,2,3$).

The amplitude and phase of each channel follow from the sublattice structure
factor,
\begin{equation}
  S^{(j)}_{ab}
  =e^{i[(\mathbf{G}_{l}+\mathbf{G}_{\alpha})\cdot\bm{\tau}_{a}
  -(\mathbf{G}'_{l'}+\mathbf{G}_{\beta})\cdot\bm{\tau}_{b}]},
\end{equation}
where $\mathbf{G}_{\alpha},\mathbf{G}_{\beta}$ are the two reciprocal vectors selected
by the delta function for the transfer $\mathbf{q}_{j}$. Near the Dirac point
$\tilde{t}(\mathbf{k}_{\alpha}+\mathbf{G}_{l}+\mathbf{G}_{\alpha})$ reduces to the
common constant $\tilde{t}(\mathbf{K}_{\alpha}^{0})$. With the graphene sublattice
positions (AA stacking before twist) $\bm{\tau}_{A}=\mathbf{0}$ and
$\bm{\tau}_{B}=(0,d)$ ($d=a/\sqrt3$), and using
$\mathbf{b}_{1}\cdot\bm{\tau}_{B}=\mathbf{b}_{2}\cdot\bm{\tau}_{B}=2\pi/3$, the
four matrix elements are
\begin{equation}
\begin{aligned}
  S^{(j)}_{AA}&=1,\qquad
  S^{(j)}_{BB}=e^{i(\mathbf{G}_{l}-\mathbf{G}'_{l'})\cdot\bm{\tau}_{B}}[1+O(\theta)],\\
  S^{(j)}_{AB}&=e^{-i\mathbf{G}'_{l'}\cdot\bm{\tau}_{B}}\,\{1,\ e^{-2\pi i/3},\ e^{2\pi i/3}\},\\
  S^{(j)}_{BA}&=e^{i\mathbf{G}_{l}\cdot\bm{\tau}_{B}}\,\{1,\ e^{2\pi i/3},\ e^{-2\pi i/3}\}.
\end{aligned}
\end{equation}
Multiplying the diagonal entries by $w_{AA}$ and the off-diagonal entries by
$w_{AB}$, each $T_{j}$ equals the BM matrix of Eq.~(\ref{eq:Tj}) up to a
sublattice gauge phase. These phases are removed by the sublattice gauge
transformation
\begin{equation}
  U_{\alpha}=\begin{pmatrix}1&0\\0&e^{i\mathbf{G}_{l}\cdot\bm{\tau}_{B}}\end{pmatrix},\qquad
  U_{\beta}=\begin{pmatrix}1&0\\0&e^{i\mathbf{G}'_{l'}\cdot\bm{\tau}_{B}}\end{pmatrix},
\end{equation}
under which $T_{j}\to U_{\alpha}^{\dagger}T_{j}U_{\beta}$ gives exactly the BM
matrices of Eq.~(\ref{eq:Tj}). Since $U_{\alpha},U_{\beta}$ are unitary the spectrum is
unchanged, and the diagonal Dirac blocks merely acquire a sublattice phase. The
folding phases are therefore pure sublattice gauge, and the framework reproduces
the BM model exactly, with the momentum conservation obtained from the cell sums
rather than assumed.

\section{Second-neighbor inter-layer coupling}
\label{app:F}
We derive the second-shell matrices of Section~4.3. The full momentum-conservation condition is $\delta_{\mathbf{q}_{l'}-\mathbf{q}_{l},\mathbf{K}_{\beta}-\mathbf{K}_{\alpha}+\mathbf{G}}$ with $\mathbf{G}$ ranging over all supercell reciprocal vectors. The first shell keeps the three values of minimal length $k_{\theta}$ (the vectors $\mathbf{q}_{1},\mathbf{q}_{2},\mathbf{q}_{3}$); the second shell keeps the next three, which are the doubled transfers $2\mathbf{q}_{1},2\mathbf{q}_{2},2\mathbf{q}_{3}$ of length $2k_{\theta}$, connecting the $\alpha$-layer Dirac point to the next-nearest $\beta$-layer images. These are inter-layer transfers, unlike the intra-layer vectors $\mathbf{q}_{i}-\mathbf{q}_{j}$ (length $\sqrt3k_{\theta}$).

Since $2\mathbf{q}_{m}$ is parallel to $\mathbf{q}_{m}$, its azimuthal angle is the same $\varphi_{m}=2\pi(m-1)/3$, but the stacking phase is squared: a stacking point at displacement $\bm{\delta}$ contributes $e^{-i\mathbf{q}_{m}\cdot\bm{\delta}}$ in the first shell and $e^{-2i\mathbf{q}_{m}\cdot\bm{\delta}}$ in the second. Hence
\begin{equation}
T_{m}^{(2)}=w_{AA}^{(2)}\sigma_{0}+w_{AB}^{(2)}(\cos 2\varphi_{m}\sigma_{x}+\sin 2\varphi_{m}\sigma_{y}),
\end{equation}
giving, with $2\varphi_{m}=\tfrac{4\pi(m-1)}{3}$, the three matrices of Eq.~(\ref{eq:Tsecond}). The diagonal $w_{AA}^{(2)}$ remains isotropic; the off-diagonal phases $e^{\pm4\pi i/3}=e^{\mp2\pi i/3}$ merely swap roles relative to the first shell, so the phase set $\{0,\pm2\pi i/3\}$ is unchanged and only the amplitudes are rescaled.

The rescaled amplitudes $w_{AA}^{(2)},w_{AB}^{(2)}$ are fixed by the same microscopic hopping evaluated at the second-shell momentum. Since $2\mathbf{q}_m\simeq 2\theta\,\hat{z}\times\mathbf{K}$ is nearly perpendicular to $\mathbf{K}$, one has $|\mathbf{K}+2\mathbf{q}_m|^2=|\mathbf{K}|^2+4k_\theta^2$, i.e.\ $|\mathbf{K}+2\mathbf{q}_m|=|\mathbf{K}|(1+\theta^2)$. With $t(q)=t_0\exp[-\alpha(qd)^\gamma]$ the ratio is
\begin{equation}
\frac{w^{(2)}}{w}\simeq\exp[-\alpha\gamma(|\mathbf{K}|d)^\gamma\theta^2]\simeq 1-\alpha\gamma(|\mathbf{K}|d)^\gamma\theta^2\simeq 0.9995,
\end{equation}
so the second shell is not suppressed: its amplitude differs from the first shell only at order $\theta^2$.

\section{Remote-band DOS convergence and the flat band at $0.67^\circ$}
\label{app:conv}

\begin{figure*}[t]
\centering
\includegraphics[width=\textwidth]{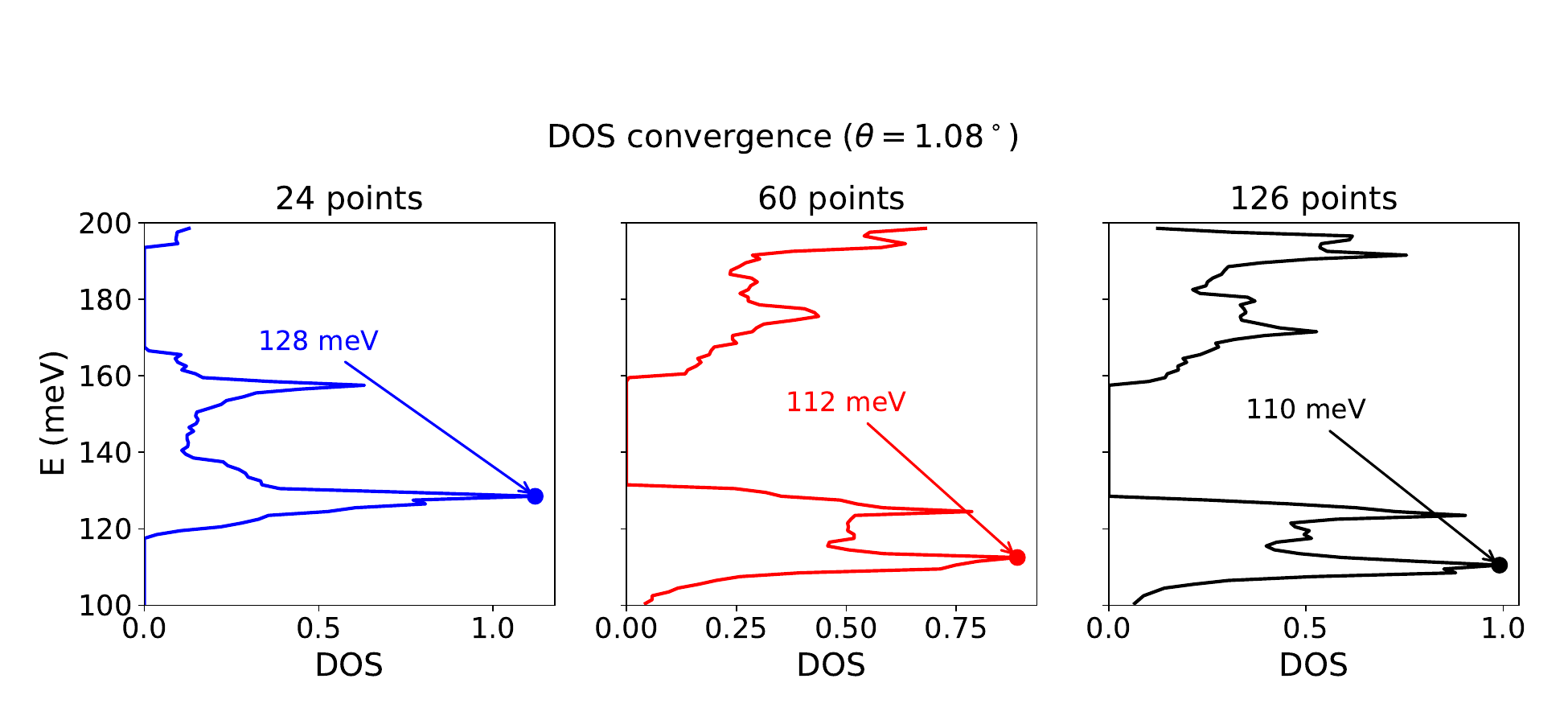}
\caption{Convergence of the remote-band DOS with the momentum-network
truncation (24, 60, and 126 points) in the $100$--$200$~meV window. The peak
moves from $\sim128$~meV to $\sim112$~meV to $\sim110$~meV, and the gap above
the peak persists in every truncation.}
\label{fig:dosconvergence}
\end{figure*}

The gap above the remote-band van Hove peak persists for every truncation of the
momentum network, so it is a robust prediction rather than a finite-size artifact.
Fig.~\ref{fig:dosconvergence} shows the remote-band DOS at $\theta=1.08^\circ$ in
the $100$--$200$~meV window for three truncations of the momentum network (24,
60, and 126 points). The peak moves from $\sim128$~meV (24 points) to
$\sim112$~meV (60 points) to $\sim110$~meV (126 points), so the 60-point
truncation is converged, and the gap above the peak persists in every truncation:
the precise peak position may shift slightly, but the existence of the gap is
model-independent.

\begin{figure}[htbp]
\centering
\includegraphics[width=\linewidth]{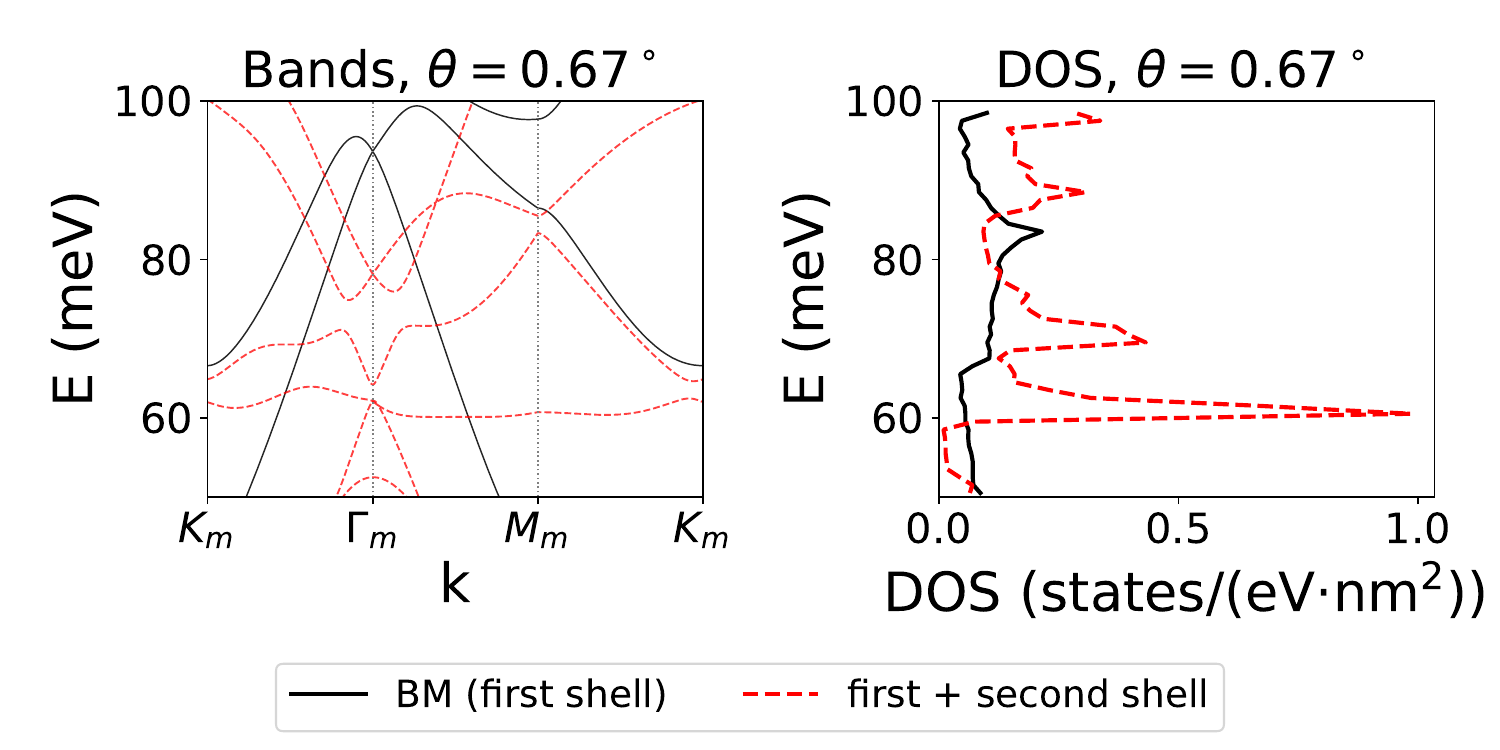}
\caption{Remote band at $\theta=0.67^\circ$ in the $50$--$100$~meV window.
Left: bands along $K_m$--$\Gamma_m$--$M_m$--$K_m$; black, first shell only
(BM); red dashed, first $+$ second shell. Right: DOS in the same window. The
second shell flattens the remote band (bandwidth $\sim4$~meV).}
\label{fig:remotedos067}
\end{figure}

The remote-band flat band is most clearly resolved at $\theta=0.67^\circ$, where
the second shell flattens the remote band into a nearly dispersionless band of
width $\sim4$~meV within the $50$--$100$~meV window (Fig.~\ref{fig:remotedos067}).

\bibliographystyle{apsrev4-2-range}
\bibliography{zw}

\end{document}